\documentclass[twocolumn,preprintnumbers,a4paper,
superscriptaddress,floatfix,twoside]{revtex4}

\usepackage{bm}
\usepackage{epsfig}
\usepackage{graphics}
\usepackage{natbib}
\usepackage{diagbox}
\usepackage{fancyh}
\usepackage[l]{floatflt}
\usepackage{epsfig}
\usepackage{amssymb}
\usepackage{latexsym}
\usepackage{times}
\usepackage{slashed}
\usepackage{upgreek}
\usepackage{amsmath}
\usepackage{amsfonts}
\usepackage{amsbsy}
\usepackage{amscd}
\usepackage{bbm}

\newcommand{\eref}[1]{(\ref{#1})}

\newcommand{\Eref}[1]{Eq.~(\ref{#1})}
\newcommand{\Erefs}[2]{Eqs.~(\ref{#1}) -- (\ref{#2})}
\newcommand{\Sref}[1]{Sec.~\ref{#1}}
\newcommand{\Srefs}[1]{Secs.~\ref{#1}}
\newcommand{\Fref}[1]{Fig.~\ref{#1}}
\newcommand{\Tref}[1]{Table~\ref{#1}}
\newcommand{\cref}[1]{Ref.~\cite{#1}}
\newcommand{\crefs}[1]{Refs.~\cite{#1}}

\newcommand{\hepph}[1]{{\ftn\tt hep-ph/#1}}

\newcommand{\astroph}[1]{{\ftn\tt astro-ph/#1}}
\newcommand{\arxiv}[1]{{\ftn\tt  arXiv:#1}}

\newcommand{\bal}{\begin{align}}
\newcommand{\eal}{\end{align}}
\newcommand{\beqs}{\begin{subequations}}
\newcommand{\eeqs}{\end{subequations}}
\newcommand{\eec}{\end{center}}
\newcommand{\bec}{\begin{center}}

\newcommand{\eem}{\end{matrix}}
\newcommand{\bem}{\begin{matrix}}
\newcommand{\eeq}{\end{equation}}
\newcommand{\beq}{\begin{equation}}
\newcommand{\ba}{\begin{array}}
\newcommand{\ea}{\end{array}}
\newcommand{\bea}{\begin{eqnarray}}
\newcommand{\eea}{\end{eqnarray}}
\newcommand{\baq}{\begin{eqnarray}}
\newcommand{\eaq}{\end{eqnarray}}\newcommand{\bel}{\begin{align}}
\newcommand{\eeel}{\begin{align}}

\newcommand{\beld}{\begin{aligned}}
\newcommand{\eeld}{\begin{aligned}}

\newcommand\eqs[2]{Eqs.~(\ref{#1}) and (\ref{#2})}

\newcommand\eqss[3]{Eqs.~(\ref{#1}), (\ref{#2}) and (\ref{#3})}

\newcommand{\ftn}{\footnotesize}

\newcommand{\TeV}{{\mbox{\rm TeV}}}

\newcommand{\GeV}{{\mbox{\rm GeV}}}
\newcommand{\eV}{{\mbox{\rm eV}}}

\newcommand{\sFref}[2]{Fig.~\ref{#1}-{\ftn\sf ({#2})}}
\newcommand{\sEref}[2]{Eq.~(\ref{#1}{\ftn\sf {#2}})}

\newcommand{\etal}{{\it et al.\/}}

\def\to{\rightarrow}

\def\llgm{\left\lgroup}
\def\rrgm{\right\rgroup}
\def\lf{\left(}
\def\rg{\right)}
\newcommand\vev[1]{\left\langle {#1} \right\rangle}
\newcommand{\Gr}{\ensuremath{\widetilde{G}}}
\newcommand{\Yb}{\ensuremath{Y_{B}}}
\newcommand{\Yg}{\ensuremath{Y_{3/2}}}
\newcommand{\Vhi}{\ensuremath{\widehat V_{\rm HI}}}
\newcommand{\Vci}{\ensuremath{\widehat V_{\rm CI}}}

\newcommand{\Hhi}{\ensuremath{\widehat H_{\rm HI}}}
\newcommand{\Hci}{\ensuremath{\widehat H_{\rm CI}}}
\newcommand{\Ohi}{\ensuremath{\Omega}}
\newcommand{\Omg}{\ensuremath{\Omega}}

\newcommand{\Whi}{\ensuremath{W_{\rm HI}}}
\newcommand{\Wci}{\ensuremath{W_{\rm CI}}}

\newcommand{\Ns}{\ensuremath{{\what N_\star}}}

\newcommand{\mP}{\ensuremath{m_{\rm P}}}

\newcommand{\Mgut}{\ensuremath{M_{\rm GUT}}}
\newcommand{\Qef}{\ensuremath{\Lambda_{\rm UV}}}

\newcommand{\lm}{\ensuremath{\lambda_\mu}}

\def\openone{\leavevmode\hbox{\small1\kern-3.8pt\normalsize1}}

\newcommand{\fr}{\ensuremath{f_{\cal R}}}
\newcommand{\frs}{\ensuremath{f_{\cal R\star}}}
\newcommand{\fn}{\ensuremath{f_n}}
\newcommand{\fns}{\ensuremath{f_{n\star}}}
\newcommand{\fm}{\ensuremath{F_{-}}}

\newcommand{\fp}{\ensuremath{F_{\cal R}}}
\newcommand{\ca}{\ensuremath{c_{\cal R}}}
\newcommand{\phc}{\ensuremath{\Phi}}
\newcommand{\phcb}{\ensuremath{\bar\Phi}}

\newcommand{\mgr}{\ensuremath{m_{3/2}}}

\newcommand{\am}{\ensuremath{{\rm a}_{3/2}}}
\newcommand{\mg}{{\ensuremath{M_{1/2}}}}

\newcommand{\Gsn}{\ensuremath{\what{\Gamma}_{\rm \dph}}}
\newcommand{\GNsn}{\ensuremath{\what{\Gamma}_{\dph\to N_i^c}}}
\newcommand{\Ghsn}{\ensuremath{\what{\Gamma}_{\dph\to H}}}
\newcommand{\Gysn}{\ensuremath{\what{\Gamma}_{\dph\to XYZ}}}
\newcommand{\msn}{\ensuremath{\what m_{\rm \dph}}}
\newcommand{\aS}{\ensuremath{{\rm a}_S}}
\newcommand{\Ald}{\ensuremath{A_\lambda}}

\newcommand{\hd}{{\ensuremath{H_d}}}
\newcommand{\hu}{{\ensuremath{H_u}}}

\newcommand{\ks}{\ensuremath{k_\star}}
\newcommand{\ns}{\ensuremath{n_{\rm s}}}
\newcommand{\as}{\ensuremath{a_{\rm s}}}
\newcommand{\As}{\ensuremath{A_{\rm s}}}
\newcommand{\rw}{\ensuremath{r_{0.002}}}

\newcommand{\rcc}{\ensuremath{\mathcal{R}}}
\newcommand{\rce}{\ensuremath{\widehat{\mathcal{R}}}}
\newcommand{\Ve}{\ensuremath{\widehat{V}}}

\newcommand{\sni}{\ensuremath{N^c_i}}
\newcommand{\ssni}{\ensuremath{\widetilde N^c_i}}

\newcommand{\mrh[1]}{\ensuremath{M_{#1N^c}}}
\newcommand{\lrh[1]}{\ensuremath{\lambda_{#1N^c}}}

\newcommand{\mD[1]}{\ensuremath{m_{#1\rm D}}}
\newcommand{\mn[1]}{\ensuremath{m_{#1\rm \nu}}}

\newcommand{\wrhn[1]}{\ensuremath{N^c_{#1}}}

\newcommand{\dphi}{\ensuremath{\what{\delta\phi}}}
\newcommand{\dph}{\ensuremath{\delta\phi}}
\newcommand{\nsu}{\ensuremath{{N_X}}}
\newcommand{\what}{\ensuremath{\widehat}}

\newcommand{\Wmu}{\ensuremath{W_{\mu}}}
\newcommand{\Wrhn}{\ensuremath{W_{\rm RHN}}}

\def\ve{\varepsilon}
\def\aal{{\bar\alpha}}
\def\bbet{{\bar\beta}}
\def\al{{\alpha}}

\def\bt{{\beta}}

\def\n{\bar{n}}

\def\th{{\theta}}
\def\thb{{\bar\theta}}

\def\thn{{\theta_{\Phi}}}

\newcommand{\Trh}{\ensuremath{T_{\rm rh}}}
\newcommand{\sg}{\ensuremath{\phi}}

\newcommand{\sgx}{\ensuremath{\phi_\star}}
\newcommand{\sgf}{\ensuremath{\phi_{\rm f}}}

\newcommand{\ld}{\ensuremath{\lambda}}
\newcommand{\ldu}{\ensuremath{\uplambda}}

\newcommand{\kp}{\ensuremath{\kappa}}

\newcommand{\se}{\ensuremath{\widehat \phi}}
\newcommand{\sex}{\ensuremath{\widehat{\phi}_\star}}
\newcommand{\sef}{\ensuremath{\widehat{\phi}_{\rm f}}}
\newcommand{\geu}{\ensuremath{\widehat g}}
\newcommand{\eph}{\ensuremath{\widehat \epsilon}}
\newcommand{\ith}{\ensuremath{\widehat \eta}}

\newcommand\mtta[4]{\mbox{
$\llgm\bem #1 &#2 \cr #3& #4\eem\rrgm$}}

\def\trns{transplanckian}
\def\Ka{K\"{a}hler potential}

\def\Kaa{K\"{a}hler~}

\def\bcp{{\sc\small Bicep2}/{\it Keck Array}}
\newcommand{\plk}{{\it Planck}}

\newcommand{\diag}{\ensuremath{{\sf diag}}}

\renewcommand{\refname}{{\bf\scshape References}}

\renewcommand{\thesubsection}{{\small\sf\Alph{subsection}}}

\renewenvironment{subequations}{%
\refstepcounter{equation}%
\setcounter{parentequation}{\value{equation}}%
  \setcounter{equation}{0}
  \ignorespaces
}{%
  \setcounter{equation}{\value{parentequation}}%
  \ignorespacesafterend
}

\begin{document}


\title{\bf\scshape Unitarity-Safe Models of Non-Minimal Inflation in Supergravity}

\author{\scshape Constantinos Pallis\\ {\it School of Technology,
Aristotle University of Thessaloniki, GR-541 12 Thessaloniki,
GREECE}\\  {\sl e-mail address: }{\ftn\tt kpallis@auth.gr}}


\begin{abstract}

\noindent {\ftn \bf\scshape Abstract:} We show that models of
chaotic inflation based on the $\phi^p$ potential and a linear
non-minimal coupling to gravity, $\fr=1+\ca\phi$, can be done
consistent with data in the context of Supergravity, retaining the
perturbative unitarity up to the Planck scale, if we employ
logarithmic \Ka s with prefactors $-p(1+n)$ or $-p(n+1)-1$, where
$-0.035\lesssim n\lesssim0.007$ for $p=2$ or $-0.0145\lesssim
n\lesssim0.006$ for $p=4$. Focusing, moreover, on a model
employing a gauge non-singlet inflaton, we show that a solution to
the $\mu$ problem of MSSM and baryogenesis via non-thermal
leptogenesis can be also accommodated.
\\ \\ {\scriptsize {\sf PACs numbers: 98.80.Cq, 11.30.Qc, 11.30.Er, 11.30.Pb, 12.60.Jv}
\hfill {\sl\bfseries Published in} {\sl Eur. Phys. J. C} {\bf 78},
no. 12, 1014 (2018) }

\end{abstract}\pagestyle{fancyplain}

\maketitle

\rhead[\fancyplain{}{ \bf \thepage}]{\fancyplain{}{\sl
Unitarity-Safe Models of nMI in SUGRA}} \lhead[\fancyplain{}{\sl
\leftmark}]{\fancyplain{}{\bf \thepage}} \cfoot{}

\section{Introduction}\label{intro}

Inflation established in the presence of a non-minimal coupling
between the inflaton $\sg$ and the Ricci scalar $\rcc$ is called
collectively \emph{non-minimal inflation} ({\sf\ftn nMI})
\cite{old, sm1, nmi, linde1, nmH, quad1, quad, talk}. Between the
numerus models, which may be proposed in this context,
\emph{universal attractor models} ({\sf\ftn UAMs}) \cite{atroest}
occupy a prominent position since they exhibit an attractor
towards an inflationary phase excellently compatible with data
\cite{plin} for $\ca\gg1$ and $\sg\leq\mP$ -- where $\mP$ is the
reduced Planck mass. UAMs consider a monomial potential of the
type
\beq \label{Vn} V_{\rm CI}(\sg)=\ld^2\sg^p/2^{p/2}\mP^{p-4}\eeq
in conjunction with a strong non-minimal coupling \cite{nmi,
atroest}
\beq \label{fr} \tilde{f}_{\cal R}(\sg)=1+\ca(\sg/\mP)^{q/2}  \eeq
with $p=q$. The emergence of an inflationary plateau in these
models can be transparently shown in the \emph{Einstein frame}
({\sf\ftn EF}) where the inflationary potential, $\what V_{\rm
attr}$, takes the form
\beq \label{VJe} \what V_{\rm attr}= {V_{\rm CI}}/{\tilde{f}_{\cal
R}^2}\simeq\ld^2\mP^{4}/\ca^2,\eeq
with the exponent in the denominator being related to the
conformal transformation employed \cite{old, sm1, nmi} to move
from the \emph{Jordan frame} ({\sf\ftn JF}) to EF.

However, due to the large $\ca$ values needed for the
establishment of nMI with $\sg\leq\mP$, the inflationary scale,
$\what V_{\rm attr}^{1/4}$, is \cite{cutoff, riotto} larger than
the \emph{Ultraviolet} ({\sf\ftn UV}) cut-off scale
\beq\Qef^{\rm attr}=\mP/\ca^{1/(q/2-1)}~~~\mbox{with}~~~
2<q\leq14/3\label{Qattr}\eeq
of the corresponding effective theory, which thereby breaks down
above it. A criticism of these results may be found in
\cref{cutof}, where background-dependent cut-off scales, well
larger than $\what V_{\rm attr}^{1/4}$, are evaluated. This
practice is rather questionable, though, preventing the
possibility of making perturbative extrapolations of the
low-energy theory. Indeed, the low-energy theory expanded around
the true low-energy vacuum should break at a scale that is
calculable within the low-energy field expansion \cite{referee}.
Therefore, the presence of $\Qef^{\rm attr}$ in \Eref{Qattr} at
lower values of the inflaton cannot be avoided, and it signals the
breakdown of the theory in that field range. Several ways have
been proposed to surpass the inconsistency above. E.g.,
incorporating new degrees of freedom at $\Qef^{\rm attr}$
\cite{gianlee}, or assuming additional interactions \cite{john},
or invoking a large inflaton \emph{vacuum expectation value}
({\sf\ftn v.e.v}) $\vev{\sg}$ \cite{R2r, nIG, igHI, gian}, or
introducing a sizable kinetic mixing in the inflaton sector which
dominates over $\tilde{f}_{\cal R}$ \cite{lee, jhep,var, nMkin,
univ}.

Here we  propose a novel solution -- applied only in the context
of \emph{Supergravity} ({\sf\ftn SUGRA}) -- to the aforementioned
problem, by exclusively considering $q=2$ in \Eref{fr} --
cf.~\cref{uvHiggs}. In this case, the canonically normalized
inflaton $\se$ is related to the initial field $\sg$ as
$\se\sim\ca\sg$ at the vacuum of the theory, in sharp contrast to
what happens for $q>2$ where $\se\simeq\sg$. As a consequence, the
small-field series of the various terms of the action expressed in
terms of $\se$, does not contain $\ca$ in the numerators,
preventing thereby the reduction of $\Qef$ below $\mP$
\cite{riotto,quad}. The same conclusion may be drawn within the JF
since no dangerous inflaton-inflaton-graviton interaction appears
\cite{quad}. Note that the importance of a scalar field with a
totally or partially linear non-minimal coupling to gravity in
unitarizing Higgs inflation within non-SUSY settings is
highlighted in \crefs{jose, uvlee}.

A permanently linear $\tilde{f}_{\cal R}$ can be reconciled with
an inflationary plateau, similar to that obtained in \Eref{VJe},
in the context of SUGRA, by suitably selecting the employed \Ka s.
Indeed, this kind of models is realized in SUGRA using logarithmic
or semilogarithmic \Ka s \cite{linde1,nmH} with the prefactor
$(-N)$ of the logarithms being related to the exponent of the
denominator in \Eref{VJe}. Therefore, by conveniently adjusting
$N$ we can achieve, in principle, a flat enough EF potential for
any $p$ in \Eref{Vn} but taking exclusively $q=2$ in \Eref{fr}. As
we show in the following, this idea works for $p\leq4$ in
\Eref{Vn} supporting nMI compatible with the present data
\cite{plin}. For $p=4$ we also show that the inflaton may be
identified with a gauge singlet or non-singlet field. In the
latter case, models of non-minimal Higgs inflation are introduced,
which may be embedded in a more complete extension of MSSM
offering a solution to the $\mu$ problem \cite{dvali} and allowing
for an explanation of \emph{baryon asymmetry of the universe}
({\ftn\sf BAU}) \cite{plcp}  via \emph{non-thermal leptogenesis}
({\ftn\sf nTL}) \cite{lept}. The resulting models employ one
parameter less than those used in \cref{univ} whereas the
gauge-symmetry-breaking scale is constrained to values well below
the MSSM unification scale contrary to what happens in
\crefs{univ,igHI}.

Below, we first -- in \Sref{sugra} -- describe the SUGRA set-up of
our models and prove that these are unitarity-conserving in
\Sref{uv}. Then, in \Sref{fhi}, we analyze the inflationary
dynamics and predictions. In \Sref{post} we concentrate on the
case of nMI driven by a Higgs field and propose a possible
post-inflationary completion. We conclude in \Sref{con}. Unless
otherwise stated, we use units where the reduced Planck scale $\mP
= 2.4\times 10^{18}~\GeV$ is set to be unity.

\section{Supergravity Framework}\label{sugra}

In Sec.~\ref{sugra1} we describe the generic formulation of our
models within SUGRA, and then we apply it for a gauge singlet and
non-singlet inflaton in \Srefs{sugra2} and \ref{sugra3}
respectively.

\subsection{\small\sf\scshape General Framework}\label{sugra1}

We focus on the part of the EF action within SUGRA related to the
complex scalars $z^\al$ -- denoted by the same superfield symbol
-- which has the form \cite{linde1}
\beqs\beq\label{Saction1}  {\sf S}=\int d^4x \sqrt{-\what{
\mathfrak{g}}}\lf-\frac{1}{2}\rce +K_{\al\bbet} \geu^{\mu\nu}D_\mu
z^\al D_\nu z^{*\bbet}-\Ve\rg\,, \eeq
where $\rce$ is the EF Ricci scalar curvature, $D_\mu$ is the
gauge covariant derivative, $K_{\al\bbet}=K_{,z^\al z^{*\bbet}}$,
and $K^{\al\bbet}K_{\bbet\gamma}=\delta^\al_{\gamma}$ --
throughout subscript of type $,z$ denotes derivation \emph{with
respect to} ({\ftn\sf w.r.t}) the field $z$. Also, $\Ve$ is the EF
SUGRA potential which can be found once we select a superpotential
$W$ in \Eref{Whi} and a \Ka\ $K$ via the formula
\beq \Ve=e^{K}\left(K^{\al\bbet}D_\al WD^*_\bbet W^*-3{\vert
W\vert^2}\right)+\frac{g^2}2 \mbox{$\sum_{\rm a}$} {\rm D}_{\rm
a}^2,\label{Vsugra} \eeq\eeqs
where $D_\al W=W_{,z^\al}+K_{,z^\al}W$ is the \Kaa\ covariant
derivative and ${\rm D}_{\rm a}=z^\al\lf T_{\rm  a}\rg_\al^\bt
K_\bt$ are the D term corresponding to a gauge group with
generators $T_{\rm a}$ and (unified) gauge coupling constant $g$.
The remaining terms in the \emph{right-hand side} ({\sf\ftn
r.h.s}) of the equation above describes contribution from the F
terms. The contribution from the D terms vanishes for a gauge
singlet inflaton and can be eliminated during nMI for a gauge
non-singlet inflaton, by identifying it with the radial part of a
conjugate pair of Higgs superfields -- see \Sref{sugra3}. In both
our scenaria, we employ a ``stabilizer" field $S$ placed at the
origin during nMI. Thanks to this arrangement, the term $3{\vert
W\vert^2}$ in $\Ve$ vanishes, avoiding thereby a possible runaway
problem, and the derivation of $\Ve$ is facilitated since the
non-vanishing terms arise from those proportional to $W_{,S}$ and
$W^*_{,S^*}$ -- see \Srefs{hiV} and \ref{hiVn} below.

Defining the frame function as
\beq\label{omgdef}
-{\Omega/N}=\exp\lf-{K}/{N}\rg\>\Rightarrow\>K=-N
\ln\lf-\Omega/N\rg\,,\eeq
where $N>0$ is a dimensionless parameter, we can obtain  -- after
a conformal transformation a long the lines of \crefs{linde1, jhep} -- the JF form of ${\sf S}$ which is  \beqs%
\begin{equation}\begin{aligned}\label{Sfinal} {\sf S}&=\int d^4x
\sqrt{-\mathfrak{g}}\lf\frac{\Omega}{2N}\rcc+{\cal
\omega}_{\al\bbet}D_\mu z^\al D^\mu z^{*\bbet}-V \right.\\
&- \left.\frac{27}{N^3}\Omega{\cal A}_\mu{\cal
A}^\mu\rg~~\mbox{with}~~{\cal
\omega}_{\al\bbet}=\Omega_{\al{\bbet}}+\frac{3-N}{N}\frac{\Omega_{\al}\Omega_{\bbet}}{\Omega}\,\cdot
\end{aligned}\end{equation}
Here we use the shorthand notation $\Omega_\al=\Omega_{,z^\al}$,
and $\Omega_\aal=\Omega_{,z^{*\aal}}$. We also set
$V=\Ve{\Omg^2}/{N^2}$ and
\beq {\cal A}_\mu =-iN \lf \Omega_\al D_\mu z^\al-\Omega_\aal
D_\mu z^{*\aal}\rg/6\Omega\,.\label{Acal}\eeq\eeqs
Although the choice $N=3$ ensures canonical kinetic terms in
\Eref{Sfinal}, $N$ may be considered in general as a free
parameter with interesting consequences not only on the
inflationary observables \cite{jhep, var, quad, roest, nIG} but
also on the consistency of the effective theory, as we show below.

\subsection{\small\sf\scshape Gauge-Singlet
Inflaton}\label{sugra2}

Below, in \Sref{hiset}, we specify the necessary ingredients
(super- and \Ka s) which allow us to implement our scenario with a
gauge-singlet inflaton. Then, in \Sref{hiV}, we outline the
derivation of the inflationary potential.

\subsubsection{\small\sf Set-up}\label{hiset}

\renewcommand{\arraystretch}{1.4}
\begin{table*}[!t]
\caption{\normalfont Mass-squared spectrum of the model defined by
\eqs{Wci}{fpm} for $K=K_1 - K_5$ along the path in \Eref{inftr}.}
\begin{ruledtabular}
\begin{tabular}{c|c|c|c|c|c|c|c}
{\sc Fields}&{\sc Eigen-} & \multicolumn{6}{c}{\sc Masses
Squared}\\\cline{3-8}
&{\sc states} && {$K=K_1$}&{$K=K_2$} & {$K=K_{3}$}&{$K=K_4$} & {$K=K_{5}$} \\
\colrule
%
4 Real &$\widehat\theta$&$\widehat m_{\theta}^2$&
\multicolumn{2}{c|}{$6(1-1/N)\Hci^2$}&\multicolumn{3}{c}{$6\Hci^2$}\\\cline{3-8}
Scalars&$\widehat s, \widehat{\bar{s}}$ &$ \widehat
m_{s}^2$&\multicolumn{2}{c|}{$6\ca\sg\Hci^2/N$}&\multicolumn{3}{c}{$6\Hci^2/N_S$}\\\hline
$2$ Weyl Spinors & $\what \psi_\pm$ & $\what m^2_{\psi\pm}$ &
\multicolumn{5}{c}{$3p\lf 1-n\ca\sg^2\rg^2\Hci^2/N\ca^2\sg^2$}
\end{tabular}\label{tab1}
\end{ruledtabular}
\end{table*}
\renewcommand{\arraystretch}{1.}

This class of models requires the utilization of two gauge singlet
chiral superfields, i.e., $z^\al=\Phi, S$, with $\Phi$ ($\al=1$)
and $S$ ($\al=2)$ being the inflaton and a ``stabilizer'' field
respectively. More specifically, we adopt the superpotential
\beq \label{Wci} W_{\rm CI}=\ld S\Phi^{p/2},\eeq
which can be uniquely determined if we impose two symmetries:
{\sf\ftn (i)} an $R$ symmetry under which $S$ and $\Phi$ have
charges $1$ and $0$; {\sf\ftn (ii)} a global $U(1)$ symmetry with
assigned charges $-1$ and $2/p$ for $S$ and $\Phi$. To obtain a
linear non-minimal coupling of $\Phi$ to gravity, though, we have
to violate the latter symmetry as regards $\phc$. Indeed, we
propose the following set of \Ka s
\beqs\bea
K_1&=&-N\ln\left(1+\ca(\fp+\fp^*)-\fm/N+F_{1S}\right),~~~\label{K1}\\
K_2&=&-N\ln\left(1+\ca(\fp+\fp^*)+F_{1S}\right)+\fm,\label{K2}\\
K_3&=&-N\ln\left(1+\ca(\fp+\fp^*)-\fm/N\right)+F_{2S},\label{K3}\\
K_4&=&-N\ln\left(1+\ca(\fp+\fp^*)\right)+\fm+F_{2S},\label{K4}\\
K_5&=&-N\ln\left(1+\ca(\fp+\fp^*)\right)+F_{3S}. \label{K5}
\eea\eeqs
Recall that $N>0$. From the involved functions
\beq \label{fpm}
\fp=\phc/\sqrt{2}~~~\mbox{and}~~~\fm=-\frac12\lf\phc-\phc^*\rg^2\eeq
the first one allows for the introduction of the linear
non-minimal coupling of $\phc$ to gravity whereas the second one
assures canonical normalization of $\Phi$ without any contribution
to the non-minimal coupling along the inflationary path --
cf.~\crefs{atroest, nMkin}. On the other hand, the functions
$F_{lS}$ with $l=1,2,3$ offer canonical normalization and safe
stabilization of $S$ during and after nMI. Their possible forms
are given in \cref{univ}. Just for definiteness, we adopt here
only their logarithmic form, i.e.,
\beqs\bea
F_{1S}&=&-\ln\left(1+|S|^2/N\right),\label{f1s}\\
F_{2S}&=&N_S\ln\left(1+|S|^2/N_S\right),\label{f2s}\\
F_{3S}&=&N_S\ln\left(1+\fm/N_S+|S|^2/N_S\right),\label{f3s}
\eea\eeqs
with $0<N_S<6$. Recall \cite{linde1,su11} that the simplest term
$|S|^2$ leads to instabilities for $K=K_1$ and $K_2$ and light
excitations for $K=K_3-K_5$. The heaviness of these modes is
required so that the observed curvature perturbation is generated
wholly by our inflaton in accordance with the lack of any
observational hint \cite{plcp} for large non-Gaussianity in the
cosmic microwave background. Note that all the proposed $K$'s
contain up to quadratic terms of the various fields. Also $\fp$
(and $\fp^*$) is exclusively included in the logarithmic part of
the $K$'s whereas $\fm$ may or may not accompany it in the
argument of the logarithm. Note finally that, although quadratic
nMI is analyzed in \crefs{quad1,atroest,quad} too, the present set
of $K$'s is examined for first time.

\subsubsection{\sf\small Inflationary Potential}\label{hiV}

Along the inflationary track determined by the constraints
\beq \label{inftr} S=\Phi-\Phi^*=0,~\mbox{or}~~s=\bar s=\th=0\eeq
if we express $\Phi$ and $S$ according to the parametrization
\beq \Phi=\:{\phi\,e^{i
\th}}/{\sqrt{2}}\>\>\>\mbox{and}\>\>\>S=\:(s +i\bar
s)/\sqrt{2}\,,\label{cannor} \eeq
the only surviving term in \Eref{Vsugra} is
\beq \label{1Vhio}\Vci=\Ve(\th=s=\bar s=0)=e^{K}K^{SS^*}\,
|W_{{\rm CI},S}|^2\,.\eeq
which, for the $K$'s in Eqs.~(\ref{K1}) -- (\ref{K5}), reads
\beq\Vci=\frac{\ld^2\sg^p}{2^{p/2}\fr^{N}}\cdot\begin{cases}
%
\fr&\mbox{for}\>\> K=K_1, K_2\\
1&\mbox{for}\>\> K=K_3 - K_5,\end{cases}\label{Vhi1}\eeq
where we define the (inflationary) frame function as
\beq \fr=-\left.{\Ohi\over N}\right|_{\rm\Eref{inftr}}=1+\ca\sg\,.
\label{from}\eeq As expected, $\fr$ coincides with $\tilde f_{\cal
R}$ in \Eref{fr} for $q=2$. This form of $\fr$ assures the
preservation of unitarity up to $\mP=1$ as explained in
\Sref{intro} and verified in \Sref{uv}. The last factor in
\Eref{Vhi1} originates from the expression of $K^{SS^*}$ for the
various $K$'s. Indeed, $K_{\al\bbet}$ along the configuration in
\Eref{inftr} takes the form
\beq \lf K_{\al\bbet}\rg=\diag\lf
\kappa+N\ca^2/2\fr^2,K_{SS^*}\rg,\label{Kab}\eeq where
\beq K_{SS^*}=\begin{cases}
%
1/\fr&\mbox{for}\>\> K=K_1, K_2\\
1&\mbox{for}\>\> K=K_3 - K_5\label{Kss}\end{cases}\eeq and
\beq\kappa=\begin{cases}
%
1/\fr&\mbox{for}\>\> K=K_1, K_3\\
1&\mbox{for}\>\> K=K_2, K_4, K_5.\label{kappa}\end{cases}\eeq
If we set
\beq \label{ndef} N= \begin{cases} p(n+1)+1 &\mbox{for}\>\> K=K_1, K_2 \\
p(n+1) &\mbox{for}\>\> K=K_3 - K_5,
\end{cases}\eeq
we arrive at a universal expression for $\Vci$ which is \beq
\label{Vhi} \Vci=
\frac{\ld^2}{2^{p/2}}\frac{\sg^p}{\fr^{p(1+n)}}\,\cdot\eeq
For $n=0$ and $p=2$, $\Vci$ reduces to $\Ve_{\rm attr}$ in
\Eref{VJe} whereas for $p>2$, $\Vci$ deviates from $\Ve_{\rm
attr}$ although it develops a similar inflationary plateau, for
$\ca\gg1$, since both numerator and denominator are dominated by a
term proportional to $\sg^p$ as in the case of UAMs.  The choice
$n=0$ is special since, for integer $p$, it yields integer $N$ in
\Erefs{K1}{K5}, i.e., $N=p+1$ for $K=K_1$ and $K_2$ or $N=p$ for
$K=K_3-K_5$. Although integer $N$'s are more friendly to string
theory -- and give observationally acceptable results as shown in
\Sref{hinum} --, non-integer $N$'s are also acceptable
\cite{roest, jhep, var, nIG, univ} and assist us to cover the
whole allowed domain of the observables. More specifically, for
$n<0$, $\Vci$ remains an increasing function of $\sg$, whereas for
$n>0$, it develops a local maximum $\Vci(\sg_{\rm max})$ where
\beq \sg_{\rm
max}=\frac1{n\ca}\lf1+\sqrt{\frac{1+n}{1+np}}\rg\,.\label{Vmax}\eeq
In a such case we are forced to assume that hilltop \cite{lofti}
nMI occurs with $\sg$ rolling from the region of the maximum down
to smaller values.

\renewcommand{\arraystretch}{1.4}
\begin{table*}[!t]
\caption{\normalfont Mass-squared spectrum of the model defined by
\eqs{Whi}{fpmH} for $K=K_1 - K_5$ along the path in
\Eref{inftrn}.}
\begin{ruledtabular}
\begin{tabular}{c|c|c|c|c|c|c|c}
{\sc Fields}&{\sc Eigen-} & \multicolumn{6}{c}{\sc Masses
Squared}\\\cline{3-8}
&{\sc states} && {$K=K_1$}&{$K=K_2$} & {$K=K_{3}$}&{$K=K_4$} & {$K=K_{5}$} \\
\colrule
%
4 Real &$\widehat\theta_{+}$&$\widehat m_{\theta+}^2$&
\multicolumn{2}{c|}{$3(1-1/N)\Hhi^2$}&\multicolumn{3}{c}{$3\Hhi^2$}\\\cline{3-8}
Scalars&$\widehat \theta_\Phi$ &$\widehat m_{ \theta_\Phi}^2$&
$M^2_{BL}$&{$M^2_{BL}+6\Hhi^2$}&\multicolumn{2}{c|}{$M^2_{BL}$}&{$M^2_{BL}+6(1+1/N_S)\Hhi^2$}\\\cline{3-8}
&$\widehat s, \widehat{\bar{s}}$ &$ \widehat
m_{s}^2$&\multicolumn{2}{c|}{$6\Hhi^2\ca\sg/N$}&\multicolumn{3}{c}{$6\Hhi^2/N_S$}\\\hline
1 Gauge Boson& $A_{BL}$ &$
M_{BL}^2$&$g^2\sg^2/\fr$&$g^2\sg^2$&$g^2\sg^2/\fr$&\multicolumn{2}{c}{$g^2\sg^2$}\\\colrule
$4$ Weyl & $\what \psi_\pm$ & $\what m^2_{\psi\pm}$ &
\multicolumn{2}{c|}{$3\lf
\ca(N-5)\sg-4\rg^2\Hhi^2/N\ca^2\sg^2$}&\multicolumn{3}{c}{$3\lf
\ca(N-4)\sg-4\rg^2\Hhi^2/N\ca^2\sg^2$}
\\\cline{3-8} Spinors &$\ldu_{BL}, \widehat\psi_{\Phi-}$&
$M_{BL}^2$&$g^2\sg^2/\fr$&$g^2\sg^2$&$g^2\sg^2/\fr$&\multicolumn{2}{c}{$g^2\sg^2$}
\end{tabular}\label{tab2}
\end{ruledtabular}
\end{table*}
\renewcommand{\arraystretch}{1.}

Defining the EF canonically normalized fields, denoted by hat, via
the relations
\beq  \label{Jg} \frac{d\se}{d\sg}=\sqrt{K_{\Phi\Phi^*}}=J,\>\>
\what{\th}= J\th\sg\>\>\mbox{and}\>\>(\what s,\what{\bar
s})=\sqrt{K_{SS^*}} {(s,\bar s)}\eeq
we can verify that the configuration in \Eref{inftr} is stable
w.r.t the excitations of the non-inflaton fields. Taking the limit
$\ca\gg1$ we find the expressions of the masses squared $\what
m^2_{z^\al}$ (with $z^\al=\th$ and $s$) arranged in \Tref{tab1},
which approach rather well the quite lengthy, exact expressions
taken into account in our numerical computation. We infer that
$\what m^2_{z^\al}\gg\Hci^2=\Vci/3$ for $1<N<6$ and $K=K_1, K_2$
or for $0<N_S<6$ and $K=K_{3} - K_5$. Therefore $m^2_{z^\al}$ are
not only positive but also heavy enough during nMI. In \Tref{tab1}
we display the masses $\what m^2_{\psi\pm}$ of the spinors
$\what\psi_\pm= (\what{\psi}_{S}\pm \what{\psi}_{\Phi})/\sqrt{2}$
too. We define $\what\psi_{S}=\sqrt{K_{SS^*}}\psi_{S}$ and
$\what\psi_{\Phi}=\sqrt{K_{\Phi\Phi^*}}\psi_{\Phi}$ where
$\psi_\Phi$ and $\psi_S$ are the Weyl spinors associated with $S$
and $\Phi$ respectively.

\subsection{\small\sf\scshape Gauge non-Singlet Inflaton}\label{sugra3}

Following the strategy of the previous section, we show below, in
\Sref{hisetn}, how we can establish a model of
unitarity-conserving nMI driven by a a gauge non-singlet inflaton
and then, in \Sref{hiVn}, we outline the derivation of the
corresponding inflationary potential.

\subsubsection{\small\sf Set-up}\label{hisetn}

In this case, as explained below \Eref{Vsugra}, we employ a pair
of left-handed chiral superfields, $\bar{\Phi}$ and $\Phi$,
oppositely charged under a gauge group, besides the stabilizer,
$S$, which is a gauge-singlet chiral superfield. Here, we take for
simplicity the group $U(1)_{B-L}$ where $B$ and $L$ denote the
baryon and lepton number respectively. We base our construction on
the superpotential \cite{susyhybrid}
\beq \Whi=\ld S\lf\bar\Phi\Phi-M^2/4\rg,\label{Whi} \eeq
where $\ld$ and $M$ are parameters which can be made positive by
field redefinitions. $\Whi$ is the most general renormalizable
superpotential consistent with a continuous R symmetry
\cite{susyhybrid} under which $S$ and $\Whi$ are equally charged
whereas $\phcb\phc$ is uncharged. To obtain nMI -- which is
actually promoted to Higgs inflation --  with a linear non-minmal
coupling to gravity we combine $\Whi$ with one of the \Ka s in
\Erefs{K1}{K5} where the functions $\fp$ and $\fm$ are now defined
as
\beq \label{fpmH}
\fp=(\phcb\phc)^{1/2}~~~\mbox{and}~~~\fm=\left|\phc-\phcb^*\right|^2\,.\eeq
Note that the proposed $K$'s respect the symmetries of $\Whi$. As
in the case of \Sref{hiset}, $\fm$ ensures that the kinetic terms
of $\bar{\Phi}$ and $\Phi$ do not enter the expression of $\fr$
along the inflationary trough -- cf. \crefs{jhep,nMkin,var,igHI}.

Comparing the resulting $K$'s with the ones used in \cref{univ} we
may notice that here $\fm$ is not accompanied by an independent
variable $c_-$ and the real function $F_+$ is here replaced by the
combination of the holomorphic function $\fp$ and its
anti-holomorphic. Therefore, the present models are more
economical since they include one parameter less. On the other
hand, the presence of unity in the argument of the logarithms
distinguishes clearly the present models from those in \cref{igHI}
where the absence of unity enforces us to invoke a large inflaton
v.e.v.

\subsubsection{\sf\small Inflationary Potential}\label{hiVn}

Employing the parameterization of $S$ in \Eref{cannor} and
expressing $\Phi$ and $\bar\Phi$ as follows
\beq\label{hpar} \Phi=\sg e^{i\th}
\cos\thn/\sqrt{2}~~\mbox{and}~~\bar\Phi=\sg e^{i\thb}
\sin\thn/\sqrt{2},\eeq
with $0\leq\thn\leq\pi/2$, we can determine a D-flat direction
from the conditions
\beq \label{inftrn} \bar
s=s=\th=\thb=0\>\>\>\mbox{and}\>\>\>\thn={\pi/4}.\eeq
Along this path the only surviving term is again given by
\Eref{1Vhio} which now reads -- cf. \Eref{Vhi1}
\beq\Vhi=\frac{\ld^2(\sg^2-M^2)^2}{16\fr^{N}}\cdot\begin{cases}
%
\fr&\mbox{for}\>\> K=K_1,K_2\\
1&\mbox{for}\>\> K=K_3-K_5,\end{cases}\label{Vhi1n}\eeq
whereas $\fr$ is again given by \Eref{from}. If we take into
account the definition of $n$ in \Eref{ndef} we end up with \beq
\label{Vhin} \Vhi=
\frac{\ld^2}{16}\frac{(\sg^2-M^2)^2}{\fr^{p(1+n)}}\,,\eeq which
can be approached rather well by \Eref{Vhi} for $p=4$, when
$M\ll1$, and $\ld^2$ replaced with $\ld^2/4$.

To specify $\se$ in the present case we note that, for all $K$'s
in \Erefs{K1}{K5} with $\fp$ and $\fm$ given in \Eref{fpmH},
$K_{\al\bbet}$ along the configuration in \Eref{inftrn} takes the
form
\beq \label{Khiab}\lf K_{\al\bbet}\rg=\diag\lf
M_\pm,K_{SS^*}\rg,\eeq where
\beq
M_\pm=\mtta{\kappa+N\ca^2/4\fr^2}{N\ca^2/4\fr^2}{N\ca^2/4\fr^2}{\kappa+N\ca^2/4\fr^2}\,\label{Mpm}
\eeq
with $\kp$ given in \Eref{kappa}. Upon diagonalization we obtain
the following eigenvalues
\beq
\label{kpm}\kp_+=\kp+N\ca^2/2\fr^{2}\>\>\>\mbox{and}\>\>\>\kp_-=\kp\,.\eeq
Inserting \eqs{hpar}{Khiab} in the second term of the r.h.s of
\Eref{Saction1} we can define the EF canonically normalized
fields, as follows
\beqs\bea \label{Je} && {d\se}/{d\sg}=J=\sqrt{\kp_+},~~\widehat
\theta_\Phi = \sg\sqrt{\kp_-}\lf\theta_\Phi-{\pi/4}\rg,~~~~~~~\\
&& \label{Je1}\widehat{\theta}_+ ={J\sg\theta_+}/\sqrt{2}
~~\mbox{and}~~\widehat{\theta}_-
=\sqrt{\kp_-}\sg\theta_-/\sqrt{2}\,,~~~~~\eea\eeqs
where $\th_{\pm}=\lf\bar\th\pm\th\rg/\sqrt{2}$ and the
normalization of $\se$, $\what s$ and $\what{\bar{s}}$ coincides
with that found in \Eref{Jg}. Note, in passing, that the spinors
$\psi_{\Phi\pm}$ associated with the superfields $\Phi$ and
$\bar\Phi$ are similarly normalized, i.e.,
$\what\psi_{\Phi\pm}=\sqrt{\kp_\pm}\psi_{\Phi\pm}$ with
$\psi_{\Phi\pm}=(\psi_\Phi\pm\psi_{\bar\Phi})/\sqrt{2}$.

To check the stability of inflationary direction in \Eref{inftrn}
w.r.t the fluctuations of the non-inflaton fields, we derive the
mass-squared spectrum of the various scalars defined in
\eqs{Je}{Je1}. Taking the limit $\ca\gg1$, we find the approximate
expressions listed in \Tref{tab2} which are rather accurate at the
horizon crossing of the pivot scale. As in case of \Tref{tab1}, we
again deduce that $\what m^2_{z^\al}\gg\Hhi^2=\Vhi/3$ for $1<N<6$
and $K=K_1, K_2$ or for $0<N_S<6$ and $K=K_{3} - K_5$. In
\Tref{tab2} we also display the masses $M_{BL}$ of the gauge boson
$A_{BL}$ and the corresponding fermions. The non-vanishing of
$M_{BL}$ signals the fact that $U(1)_{B-L}$ is broken during nMI
and so no cosmic string are produced at its end. Finally, the
unspecified eigenstates $\what \psi_\pm$ are defined as $\what
\psi_\pm=(\what{\psi}_{\Phi+}\pm \what{\psi}_{S})/\sqrt{2}$ -- cf.
\Tref{tab1}.

\section{Effective Cut-Off Scale}\label{uv}

The motivation of our proposal originates from the fact that $\fr$
in \Eref{from} assures that the perturbative unitarity is retained
up to $\mP$ although that the attainment of nMI for $\sg\leq\mP$
requires large $\ca$'s -- as expected from the UAMs
\cite{nmi,quad,quad1} and verified in \Sref{fhi} below. To show
that this achievement is valid, we extract below the UV cut-off
scale, $\Qef$, expanding the action in the JF -- see \Sref{uv1} --
and in the EF -- see \Sref{uv2}. Throughout this section, we find
it convenient to restore $\mP$ in the formulas. We concentrate,
also, on the versions of our model with gauge singlet inflaton.
However, this analysis covers also the case of a gauge non-singlet
inflaton for $M\ll\mP$, $p=4$ and $\ld^2$ replaced by $\ld^2/4$.

Although the expansions about $\vev{\phi}=0$, presented below, are
not valid \cite{cutof} during nMI, we consider $\Qef$ extracted
this way as the overall cut-off scale of the theory for two
reasons: {\ftn\sf (i)} the reheating phase -- realized via
oscillations about $\vev{\sg}$ -- is an unavoidable stage of the
inflationary dynamics; {\ftn\sf (ii)} the result is within the
range of validity of the low-energy theory and so this can be
perturbatively extrapolated up to $\Qef$.

\subsection{\sf\ftn Jordan Frame Computation}\label{uv1}

Thanks to the special dependence of $\fr$ on $\phi$ in
\Eref{from}, there is no interaction between the excitation of
$\phi$ about $\vev{\phi}=0$, $\dph$, and the graviton,
$h^{\mu\nu}$ which can jeopardize the validity of perturbative
unitarity. To show this, we first expand $g_{\mu\nu}$ about the
flat spacetime metric $\eta_{\mu\nu}$ and the inflaton $\phi$
about its v.e.v,
\beq \label{excit}
g_{\mu\nu}\simeq\eta_{\mu\nu}+h_{\mu\nu}/\mP\>\>\>\mbox{and}\>\>\>\phi=0
+\dph\,.\eeq
Retaining only the terms with up to two space-time derivatives of
the excitations, the part of the lagrangian corresponding to the
two first terms in the r.h.s of \Eref{Sfinal} takes the form
\cite{cutof, nIG}
\begin{align} \delta{\cal L}&=-{\vev{\fr}\over8}{G}_{\rm EH}\lf h^{\mu\nu}\rg
+\frac12\vev{f_{\rm K}}\partial_\mu
\dph\partial^\mu\dph\nonumber\\&+\frac{\mP}{2}G_{\cal R}\lf
h^{\mu\nu}\rg\lf
\vev{f_{\cal R,\phi}}\dph+\sum_{\ell=2}^\infty\frac1{\ell!} \vev{\frac{d^\ell f_{\cal R}}{d\sg^\ell}}\dph^\ell\rg\nonumber\\
&=-{1\over8}G_{\rm EH}\lf \bar h^{\mu\nu}\rg+ \frac12\partial_\mu
\overline\dph\partial^\mu\overline\dph,\label{L2}\end{align}
where the functions $G_{\rm EH}$ and $G_{\cal R}$ are identical to
$F_{\rm EH}$ and $F_{\cal R}$ defined in \cref{nIG} -- $F_{\cal
R}$ should not to be confused with that given in \eqs{fpm}{fpmH}.
From \Eref{Sfinal} we compute $f_{\rm
K}=\omega_{\Phi\Phi^*}=1/\kp+(N-3)\ca^2/2\fr$. For $\vev{\sg}=0$
we get $\vev{f_{\rm K}}=1+(N-3)\ca^2/2$, $\vev{\fr}=1$ and
$\vev{f_{\cal R,\phi}}=\ca/\mP$. The JF canonically normalized
fields $\bar h_{\mu\nu}$ and $\overline\dph$, which diagonalise
the quadratic part of the form above, are defined as
\bel
\overline\dph\simeq\sqrt{\frac{N}{2}}\ca\dph\>\>\>\mbox{and}\>\>\>
\bar h_{\mu\nu}= h_{\mu\nu}+\ca\eta_{\mu\nu}\dph\,. \label{Jcan}
\end{align}

The UV behavior of the scattering amplitudes is determined by the
operators with dimension higher than four which arise from the
last term in the second line of \Eref{L2}. The resulting
interactions are proportional to the quantity
$\overline\dph^\ell\Box\bar h/\mP^{\ell-1}$ with
$\Box=\partial^\mu\partial_\mu$ and $\bar h = \bar h_\mu^\mu$ --
cf.~\cref{cutof, R2r, nIG, uvHiggs}. Given that $d^\ell f_{\cal
R}/d\sg^\ell$ vanishes for $\ell>2$, though, no interaction of
that type appears and so the theory does not face any problem with
the perturbative unitarity.

\subsection{\sf\ftn Einstein Frame Computation}\label{uv2}

Alternatively, $\Qef$ can be determined in EF, analyzing the
small-field behavior of our models following \cref{riotto}. We
focus first on the second term in the r.h.s of \eref{Saction1} for
$\mu=\nu=0$ and $K_{\phc\phc^*}=J^2$ given in \Eref{Jg}. Expanding
it about $\vev{\phi}=0$, we arrive at
\bea\nonumber
J^2\dot\phi^2&=&\lf1+\frac12N\ca^2-N\ca^3\frac{\dph}{\mP}+\frac32N\ca^4\frac{\dph^2}{\mP^2}\right.\\
&&\left.-2N\ca^5\frac{\dph^3}{\mP^3}+\cdots\rg\dot{\dph}^2,\label{J1exp}
\eea
where $\dph$ is defined in \Eref{excit} and we set
$\dot\phi=\dot{\dph}$. If the EF (canonically normalized)
inflaton, $\dphi$, was identical to $\dph$ at the vacuum of the
theory -- as happens for the UAMs \cite{cutoff, riotto} with $q$
in the range of \Eref{Qattr} -- $\Qef$ would have been equal to
$\mP/\ca^3$ since this is the lowest of the denominators above for
$\ca\gg1$. However, here we have $\dphi=\vev{J}\dph\gg\dph$ since
\beq \vev{J}=\sqrt{1+N\ca^2/2}\simeq\ca\sqrt{N/2}.\label{vevJ}\eeq
Therefore, expressing \Eref{J1exp} in terms of $\dphi$ and
$\dot{\dphi}=\vev{J}\dot{\dph}$, we see that the $\ca$'s are
cancelled out and we end up with the result
\beqs\beq\begin{aligned} \label{Jexp}
J^2\dot\phi^2=\lf1-2\frac{\sqrt{2}}{\sqrt{N}}\frac{\dphi}{\mP}+
\frac{6}{N}\frac{\dphi^2}{\mP^2}-
{8}\frac{\sqrt{2}}{N^\frac32}\frac{\dphi^3}{\mP^3}+\cdots\rg\dot{\dphi}^2
\end{aligned}\eeq
where we neglect terms suppressed by powers of $\ca$ in the
denominator. This expression is valid for both cases in \Eref{Kab}
and agrees with that in \cref{quad} for $N=3(1+n)$. Expanding
similarly $\Vci$, see \Eref{Vhi}, in terms of $\dphi$ we have
\bea\nonumber
\Vci&=&\ld^2\mP^4\lf\frac{\dphi}{\sqrt{N}\ca}\rg^{p}\lf1-\sqrt{\frac{2}{N}}(1+n)p\frac{\dphi}{\mP}\ +\right.\\
&&\frac{p}{N}\big(1+p(1+n)\big)(1+n)\frac{\dphi^2}{\mP^2}+\cdots\Bigg).
\label{Vexp}\eea\eeqs
Since no numerator proportional to $\ca$ arises in
\eqs{Jexp}{Vexp}, we conclude that $\Qef=\mP$. As a consequence,
our inflationary model is not sensitive to the UV completion of
the theory, provided that $\Vci^{1/4}(\sg)\ll\mP$ with
$\sg\leq\mP$. These prerequisites are readily fulfilled as we see
in \Sref{hinum}.

\section{Inflation Analysis}\label{fhi}

In \Srefs{hiobs} and \ref{hinum} below we examine
semi-analytically and numerically respectively, if $\Vci$ in
\Eref{Vhi} may be consistent with a number of observational
constraints. The analysis can be easily adapted to the case of
$\Vhi$ in \Eref{Vhin} performing the replacements mentioned in
\Sref{uv}.

\subsection{\sc\small\sffamily Semi-Analytic Results}\label{hiobs}

The period of slow-roll nMI is determined in the EF by the
condition -- see, e.g., \cref{review}:
\beqs\beq{\ftn\sf
max}\{\eph(\phi),|\ith(\phi)|\}\leq1,\label{srcon}\eeq where the
slow-roll parameters $\eph$ and $\ith$ read
\beq\label{sr}\widehat\epsilon= \left({\Ve_{\rm
CI,\se}/\sqrt{2}\Ve_{\rm CI}}\right)^2
\>\>\>\mbox{and}\>\>\>\>\>\widehat\eta={\Ve_{\rm
CI,\se\se}/\Ve_{\rm CI}} \eeq\eeqs
and can be derived employing $J\simeq\sqrt{N/2\sg^2}$ in \Eref{Jg}
without express explicitly $\Vci$ in terms of $\se$. Since $J$ for
$K=K_1, K_3$ deviates slightly from that for $K=K_2,K_4,K_5$ --
see \Eref{kappa} -- we have a discrimination as regards the
expressions of $\eph$ and $\ith$ in these two cases. Indeed, our
results are
\begin{equation}
\label{eph}  \eph\simeq p^2\frac{\fn^2}{\ca^2\sg^2}\begin{cases}
1/N&\mbox{for}~~K=K_1,K_3\,,
\\ 1/(N+2\sg^2)&\mbox{for}~~K=K_2,K_4,K_5\,,
\end{cases}
\end{equation}
where $\fn=1-n\ca\sg$. Similarly, we obtain
\begin{equation}
\label{ith}  \frac{\ith}{2p}\simeq \begin{cases}
(p\fn^2-\fr)/{N\ca^2\sg^2}&\hspace*{-3.3cm}\mbox{for}~~K=K_1,K_3\,,
\\ \lf{N\ca\sg\fn+(N+2\sg^2)(\ca\sg(\fn-1)+p\fn^2)}\rg/ \\ {\ca^2\sg^2(N+2\sg^2)}&
\hspace*{-3.3cm}\mbox{for}~~K=K_2,K_4,K_5\,.
\end{cases}
\end{equation}
For any of the $K$'s above we can numerically verify that
\Eref{srcon} is saturated for $\sg=\sgf$, which is found from the
condition
\beq \eph\lf\sgf\rg\simeq1~~\Rightarrow~~\sgf\simeq
\frac{p}{\ca}\frac{\sqrt{N}-pn}{N-n^2p^2}\,\cdot \label{sgf}\eeq
Apart from irrelevant constant prefactors, the formulas above for
$n=0$ and $K=K_1, K_3$ reduce to the ones obtained for quadratic
nMI -- cf.~\crefs{quad1,talk}.

The number of e-foldings, $\Ns$, that the pivot scale
$\ks=0.05/{\rm Mpc}$ experiences during nMI and the amplitude
$\As$ of the power spectrum of the curvature perturbations
generated by $\sg$ can be computed using the standard formulae
\begin{equation}
\label{Nhi}  \mbox{\sf\small (a)}~~\Ns=\int_{\sef}^{\sex}
d\se\frac{\Vci}{\Ve_{\rm CI,\se}}~~~\mbox{and}~~~\mbox{\sf\small
(b)}~~\As=\frac{1}{12\pi^2} \; \frac{\Ve_{\rm
CI}^{3}(\sex)}{\Ve_{\rm CI,\se}^2(\sex)},\eeq
where $\sgx~[\sex]$ is the value of $\sg~[\se]$ when $\ks$ crosses
the inflationary horizon. Taking into account $\sgx\gg\sgf$, we
can derive $\Ns$. We single out the following cases:

\begin{itemize}

\item For $n=0$ and any $K$ in \Erefs{K1}{K5}, we obtain
\beqs\begin{equation} \label{Nhi1}
\Ns=\frac{N\ca}{2p}\sgx~~\Rightarrow~~\sgx\simeq\frac{2p\Ns}{N\ca}\,.
\end{equation}
Note that for $n\neq0$ the formulas below for $\Ns$ cannot be
reduced to the previous one.

\item For $n\neq0$ and $K=K_1, K_3$, we obtain
\begin{equation}
\label{Nhi2}
\Ns=-\frac{N\ln\fns}{2n(1+n)p}~~\Rightarrow~~\sgx\simeq\frac{1-e_n}{n\ca},
\end{equation}
where $e_n=e^{-2pn(n+1)\Ns/N}$ and $\fns=\fn(\sgx)$.

\item For $n<0$ and $K=K_2, K_4, K_5$, we obtain
\begin{equation}
\label{Nhi3}
\Ns=-\frac{N\ln\fns}{2n(1+n)p}-\frac{\sgx^2}{2np}~~\Rightarrow~~\sgx\simeq\sqrt\frac{N}{2(n+1)}W_0(y),
\end{equation}
where $W_k$ is the Lambert or product logarithmic function
\cite{wolfram} with $k=0$ and \beq y=2(1+n)e_n^2/Nn^2\ca^2.\eeq

\item For $n>0$ and $K=K_2, K_4, K_5$, we obtain
\begin{equation}
\label{Nhi4}
\Ns=-\frac{2+n^2\ca^2N}{2n\ca^2n^3(1+n)p}\ln\fns-\frac{3+\fns}{n^2p\ca}\sgx\,.
\end{equation}\eeqs
Here we are not able to solve the equation above w.r.t $\sgx$. As
a consequence, it is not doable to find an analytical expression
for $\sgx$ and the inflationary observables -- see below.
Therefore, in this portion of parameter space, our last resort is
the numerical computation, whose the results are presented in
\Sref{hinum}.
\end{itemize}

In all cases above, there is a lower bound on $\ca$, above which
$\sgx\leq1$, as in the original UAMs \cite{atroest}. E.g., for
$n=0$
\begin{equation}
\label{camin}\sgx\leq1~~\Rightarrow~~\ca\geq{2p\Ns}/{N}\gg1\,,
\end{equation}
with $\Ns\simeq(52-59)$ for $p=2-4$. As shown in \Sref{uv}, these
large $\ca$'s do not disturb the validity of perturbative
unitarity up to $\mP$. On the other hand, the EF field, $\se$ may
be \trns, since integrating the first equation in \Eref{Jg} with
$J$ given below \Eref{sr}, we find
\beq \se\simeq\se_{\rm c}+\sqrt{N/2}\ln\sg,\label{se1}\eeq
with $\se_{\rm c}$ being a constant of integration. If we set,
e.g., $\se_{\rm c}=0$ and $\sg\simeq(0.001-1)$, then $|\se|>1$.
Despite this fact, our proposal is stabilized against corrections
from higher order terms in $\Wci$ and/or $K$'s in Eqs.~(\ref{Wci})
and (\ref{K1}) -- (\ref{K5}), since these terms are exclusively
expressed as functions of the initial field $\Phi$ and remain
harmless for $\phi=\sqrt{2}|\phc|\leq1$ -- cf. \cref{atroest}.

From \sEref{Nhi}{b} we can derive a relation between $\ld$ and
$\ca^{p/2}$ for fixed $n$, in sharp contrast to UAMs where the
same condition implies a relation between $\ld$ and $\ca$
\cite{nmH,quad1,talk}. Given that $\ca$ assumes large values, we
expect that $\ld$ increases with $p$ and rapidly (for $p\geq5$ as
we find numerically) violates the perturbative bound
$\ld\leq2\sqrt{\pi}\simeq3.5$. In particular, our results can be
cast as following:

\begin{itemize}

\item For $n=0$ and any $K$ in \Erefs{K1}{K5}, we obtain
\beqs\begin{equation} \label{lan1}
\ld=2^{\frac12-\frac{p}4}\pi\sqrt{3N\As}\ca^{p/2}/\Ns\,.
\end{equation}

\item For $n\neq0$ and $K=K_1, K_3$, we obtain
\begin{equation}
\label{lan2} \ld=2^{\frac32+\frac{p}{4}}p\pi\sqrt{3\As}(1-n\frs)
\frs^{\frac{np}{2}-1}\ca^{{p}/{2}}/{N^{1/2}},
\end{equation}
where $\frs=\fr(\sgx)$. Taking into account that $n\ll1$ and
$\frs$ is almost proportional to $\ca$ for large $\ca$, we can
easily convince ourselves that the output above implies that
$\ld/\ca^{p/2}$ remains constant for fixed $n$.

\item For $n<0$ and $K=K_2, K_4, K_5$, we obtain
\begin{equation}
\label{lan3}
\ld=2^{\frac32+\frac{p}{4}}p\pi\sqrt{3\As}\frac{(1-n\frs)
\frs^{\frac{np}{2}-1}\ca^{1+\frac{p}{2}}}{(2\frs^2+N\ca^2)^{1/2}},
\end{equation}\eeqs
from which we can again verify that the approximate
proportionality of $\ld$ on $\ca^{p/2}$ holds.

\end{itemize}

The remaining inflationary observables -- i.e., the (scalar)
spectral index $\ns$, its running $\as$, and the scalar-to-tensor
ratio $r$ -- are found from the relations \cite{review}
\beqs\bea \label{ns} && \ns=\: 1-6\eph_\star\ +\
2\ith_\star,~~r=16\eph_\star, \\ \label{as} && \as
=\:2\left(4\ith_\star^2-(n_{\rm
s}-1)^2\right)/3-2\widehat\xi_\star, \eea\eeqs
where the variables with subscript $\star$ are evaluated at
$\sg=\sgx$ and $\widehat\xi={\Ve_{\rm CI,\widehat\phi} \Ve_{\rm
CI,\widehat\phi\widehat\phi\widehat\phi}/\Ve^2_{\rm CI}}$.
Inserting $\sgx$ from \eqss{Nhi1}{Nhi2}{Nhi3} into \Eref{sr} and
then into equations above we can obtain some analytical estimates.
Namely:

\begin{figure*}[!t]
\includegraphics[width=60mm,angle=-90]{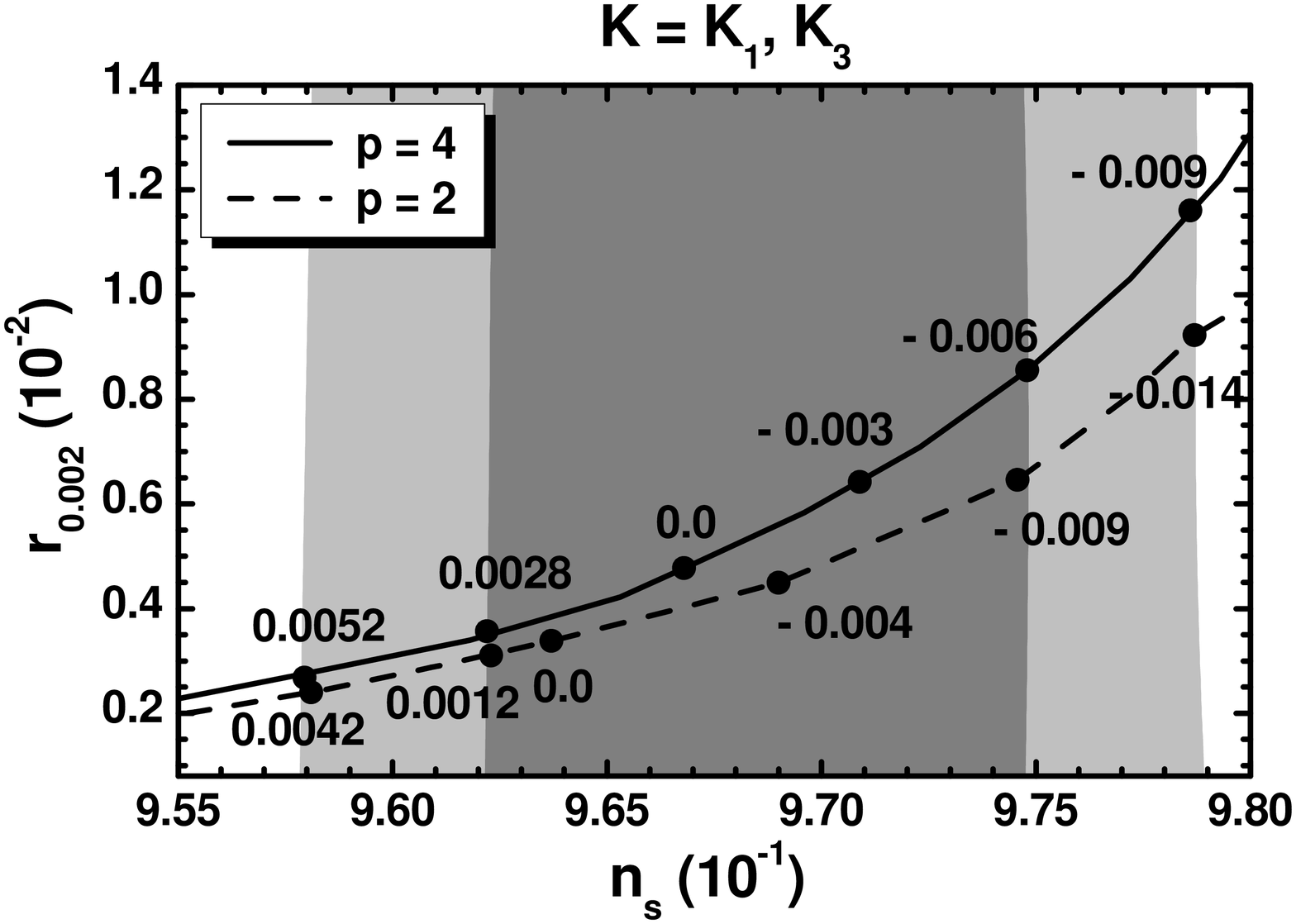}
\includegraphics[width=60mm,angle=-90]{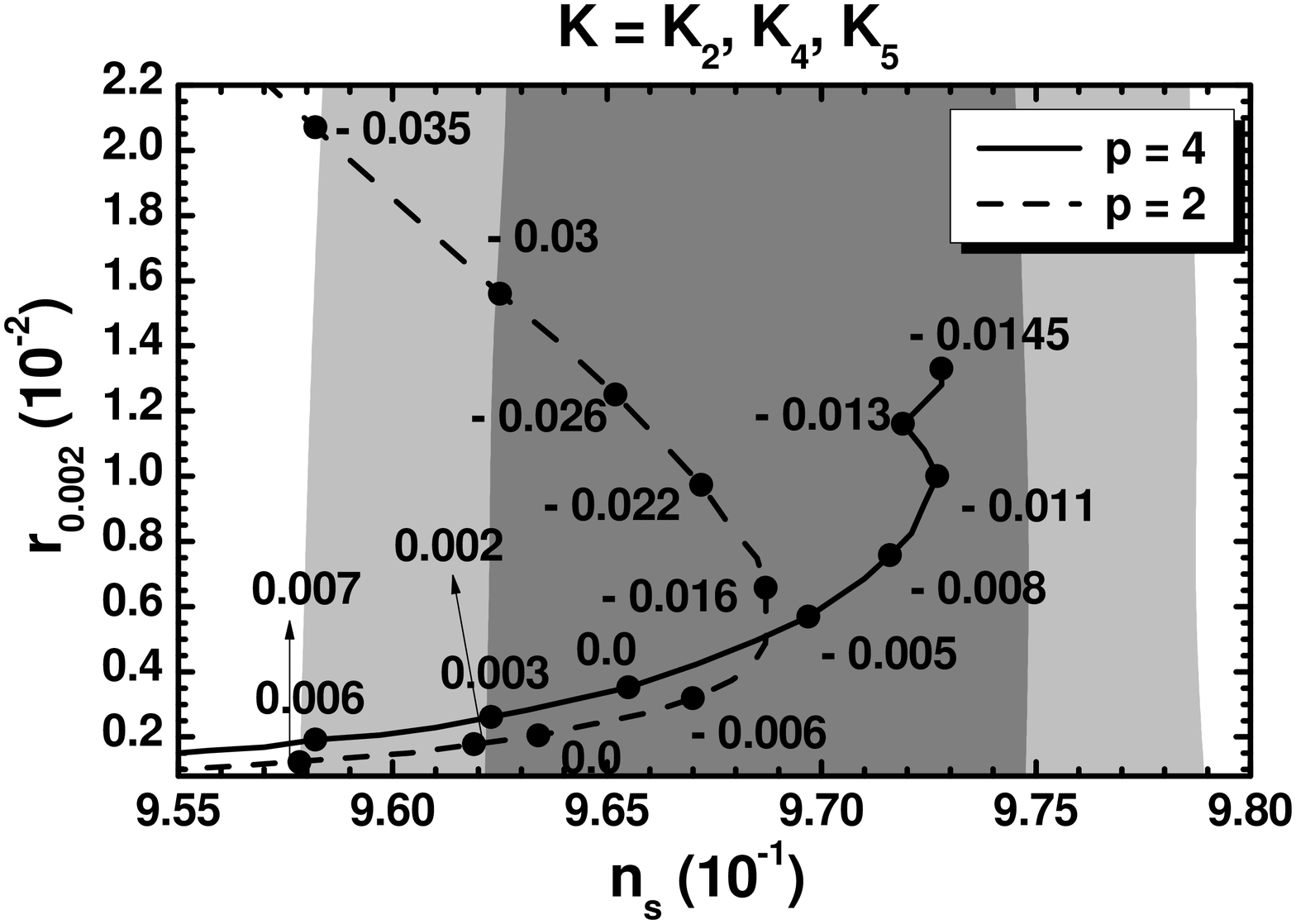}
\caption{\sl Allowed curves in the $\ns-\rw$ plane for $p=2$
(dashed lines) and $p=4$ (solid lines), $K=K_1$ and $K_3$ (left
plot) or $K=K_2, K_4$ and $K_5$ (right plot) with the various
$n$'s indicated on the lines. The marginalized joint $68\%$
[$95\%$] regions from \plk, BAO and BK14 data \cite{gwsnew} are
depicted by the dark [light] shaded contours. For $K=K_2, K_4$ and
$K_5$ the results are obtained for $\sgx\simeq1$. }\label{fig1}
\end{figure*}

\begin{itemize}

\item For $K=K_1$ and $K_3$ we end up with a unified result \beqs
\bea\label{ns1}\nonumber\ns&\simeq&1 - \frac{4p}{N\frs}
-\frac{2p^2}{N}\lf n-\frac{1}{\frs}\rg^2\\
&\simeq&1-\frac2\Ns+\frac{np}\Ns-\frac{2np}{N};\\
r&\simeq&\frac{16p^2}{N}\lf n-\frac{1}{\frs}\rg^2\simeq\frac{4N}{\Ns^2}-\frac{8np}{\Ns};\label{rs1}\\
\nonumber\as&\simeq&8p^2(n\frs-1)\frac{1+(1+n)p+\frs(1-np)}{N^2\frs^3}\\&\simeq&-\frac{2}{\Ns^2}+n\frac{1+p}{\Ns^2}\,.
\label{as1}\eea \eeqs For $n=0$ the above results are also valid
for $K=K_2, K_4$ and $K_5$ and yield observables identical with
those obtained within UAMs \cite{linde1,nmH,quad1}.

\item  For $n<0$ and $K=K_2, K_4$ and $K_5$ we arrive at the
following results \beqs\bea\nonumber&&\ns\simeq 1 +
\frac{2p\ca^2}{\frs^2(2\frs^2 +N\ca^2)^2}\Big(4 \frs^3 (n\frs -2) \\
&&-2N \ca^2\frs - (n\frs -1)^2 (2\frs^2
+N\ca^2) p\Big);~~~\label{ns2}\\
&&r\simeq\frac{16p^2}{N+2\frs^2/\ca^2}\lf
n-\frac{1}{\frs}\rg^2\label{rs2}\eea\eeqs with negligibly small
$\as$, as we find out numerically. Contrary to our previous
results, here a $\ca$ dependence arises which complicates somehow
the investigation of these models -- see \Sref{hinum}.

\end{itemize}

\subsection{\sc\sf Numerical Results}\label{hinum}

The conclusions above can be verified and extended for any $n$
numerically. In particular, we confront the quantities in
\Eref{Nhi} with the observational requirements \cite{plcp}
\beqs\bea  \nonumber&&\hspace*{-2.3mm}\Ns
\simeq61.3+\frac{1-3w_{\rm rh}}{12(1+w_{\rm
rh})}\ln\frac{\pi^2g_{\rm rh*}\Trh^4}{30\Vci(\sgf)\fr(\sgf)^2}+\\
&&\hspace*{-2.8mm} \frac12\ln\lf{\Vci(\sgx)\fr(\sgx)\over g_{\rm
rh*}^{1/6}\Vci(\sgf)^{1/2}}\rg~~
\mbox{and}~~\As\simeq2.141\cdot10^{-9},~~~~~~\label{Prob}\eea\eeqs
where we assume that nMI is followed in turn by a oscillatory
phase, with mean equation-of-state parameter $w_{\rm rh}\simeq0$
or $1/3$ for $p=2$ or $4$ respectively \cite{plcp}, radiation and
matter domination. Also $\Trh$ is the reheat temperature after
nMI, with energy-density effective number of degrees of freedom
$g_{\rm rh*}=228.75$ which corresponds to the MSSM spectrum -- see
\Sref{lept0}.

Enforcing \Eref{Prob} we can restrict $\ld/\ca^{p/2}$ and $\sgx$
and compute the model predictions via \eqs{ns}{as}, for any
selected $p$ and $n$. The outputs, encoded as lines in the
$\ns-\rw$ plane, are compared against the observational data
\cite{plcp,gwsnew} in \Fref{fig1} for $K=K_1$ and $K_3$ (left
plot) or $K=K_2, K_4$ and $K_5$ (right plot) -- here
$\rw=16\eph(\sg_{0.002})$ where $\sg_{0.002}$ is the value of
$\sg$ when the scale $k=0.002/{\rm Mpc}$, which undergoes $\what
N_{0.002}=(\Ns+3.22)$ e-foldings during nMI, crosses the horizon
of nMI. We draw dashed [solid] lines for $p=2$ [$p=4$] and show
the variation of $n$ along each line. We take into account the
data from \plk\ and \emph{Baryon Acoustic Oscillations} ({\sf\ftn
BAO}) and the {\sf\ftn BK14} data taken by the \bcp\ CMB
polarization experiments up to and including the 2014 observing
season. Fitting the data above \cite{plin,gwsnew} with
$\Lambda$CDM$+r$ model we obtain the marginalized joint $68\%$
[$95\%$] regions depicted by the dark [light] shaded contours in
\Fref{fig1}. Approximately we get
\begin{equation}  \label{nswmap}
\mbox{\ftn\sf
(a)}\>\>\ns=0.968\pm0.009\>\>\>~\mbox{and}\>\>\>~\mbox{\ftn\sf
(b)}\>\>r\leq0.07,
\end{equation}
at 95$\%$ \emph{confidence level} ({\sf\ftn c.l.}) with
$|\as|\ll0.01$.

From the left plot of \Fref{fig1} we observe that the whole
observationally favored range of $\ns$ is covered varying $n$
which, though, remains close to zero signalizing an amount of
tuning. In accordance with \eqs{ns1}{rs1}, we find the allowed
ranges
\beqs\beq\label{res21}
0.42\gtrsim{n}/{0.01}\gtrsim-1.4\>\>\>\mbox{and}\>\>\> 3\lesssim
{r}/{10^{-3}}\lesssim9.8\eeq for $p=2$, whereas for $p=4$ we have
\beq \label{res41} 5.2\gtrsim
{n/0.001}\gtrsim-9\>\>\>\mbox{and}\>\>\> 3\lesssim
{r/10^{-3}}\lesssim11\,. \eeq\eeqs
As $n$ varies in its allowed ranges above, we obtain
\beq 1.3\lesssim{10^{5}\ld/\ca}\lesssim2.3
\>\>\mbox{or}\>\>2.1\lesssim{10^5}\ld/\ca^2\lesssim3.7,
\label{res241}\eeq
for $p=2$ or $p=4$ respectively in agreement with
\eqs{lan1}{lan2}. If we take $n=0$, we find $\ns=0.963$,
$\as\simeq-6.7\cdot10^{-4}$ and $r=0.004$ for $p=2$ demanding
$\Ns\simeq53$, whereas for $p=4$ we get $\ns=0.967$,
$\as\simeq-5.6\cdot10^{-4}$ and $r=0.005$ requiring
$\Ns\simeq58.6$. Therefore, for integer prefactors of the
logarithms in \eqs{K1}{K3}, $\ns$ converges towards its central
value in \Eref{nswmap} and practically coincides with the
prediction of the UAMs \cite{atroest,nmi,nmH}. The results for
$p=4$ are rather close to those achieved in \cref{igHI} for
$K=K_{1\cal R}$ and $K_{2\cal R}$ and approach the ones found in
\cref{univ} for the largest possible value of the parameter
$r_\pm$.

Fixing, in addition, $n=0$ and $K=K_1$, $\sgx=1$ -- i.e. confining
the corresponding $\ca$ and $\ld$ values to their lowest possible
values enforcing \Eref{lan1} -- we illustrate in \Fref{fig3} the
structure of $\Vci$ as a function of $\sg$ for $p=2$ (light gray
line) or $p=4$ (gray line). More specifically, we find
$\ld=1.173\cdot10^{-3}$ or $0.257$ and $\ca=75$ or $99$ for $p=2$
or $p=4$ respectively. We see that in both cases $\Vci$ develops a
plateau with magnitude $10^{-10}$ which is similar to that
obtained in Starobinsky inflation \cite{su11,R2r} but one order of
lower than that obtained from the models analyzed in
\crefs{univ,var} where $r$ is a little more enhanced. Contrary to
Starobinsky inflation, though, $\Vci$ is well defined at the
origin as happens within the original UAMs and those in
\crefs{univ,var}. Obviously, the requirement -- mentioned in
\Sref{uv} -- $\Vci^{1/4}\ll1$ dictated from the validity of the
effective theory is readily fulfilled.

\begin{figure}[!t]
\centering
\includegraphics[width=60mm,angle=-90]{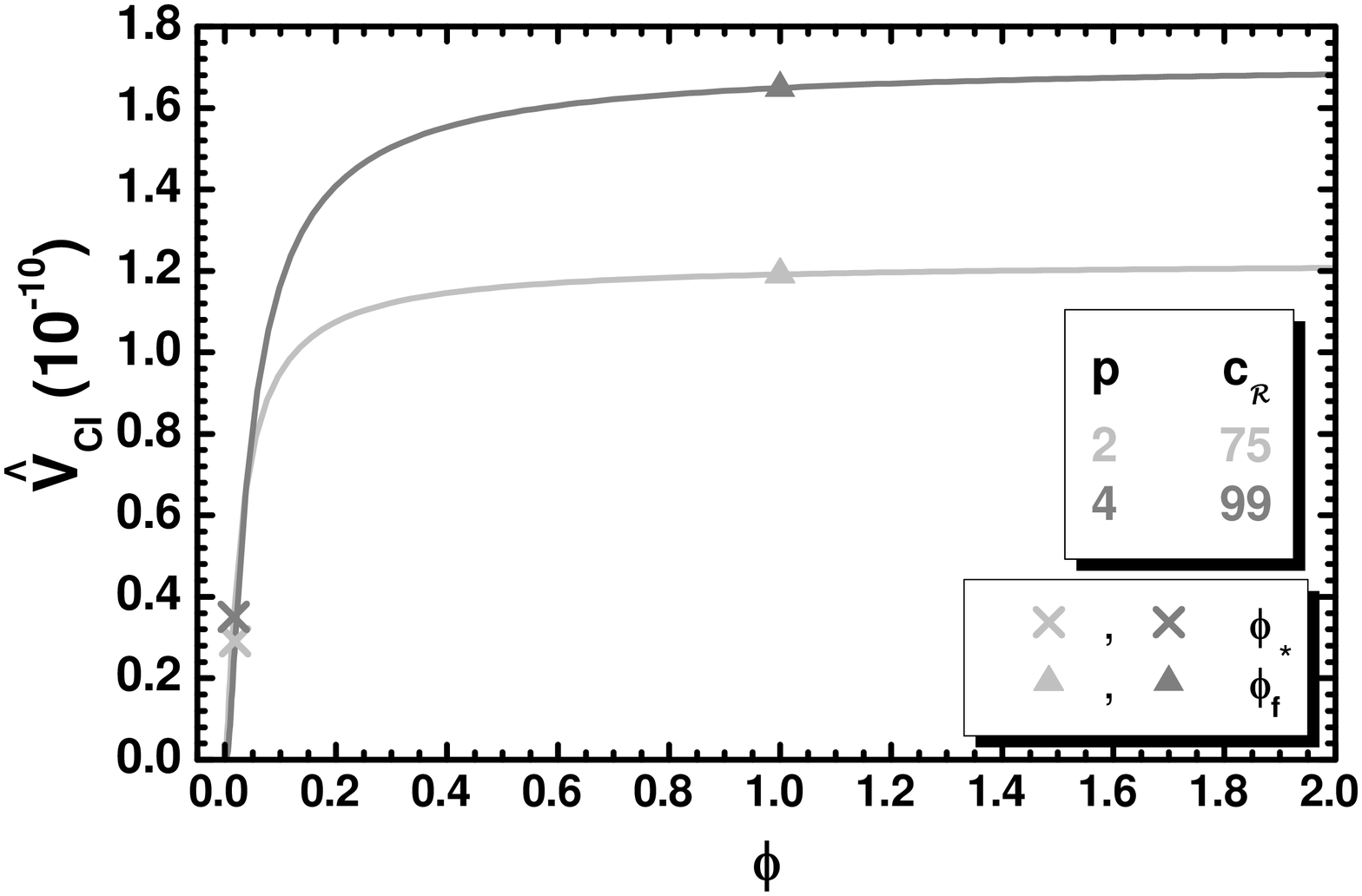}\hspace{-.1mm}
\caption{\sl \small The inflationary potential $\Vci$ as a
function of $\sg$ for $K=K_1$, $\sgx=1, n=0$ and $p=2$ (light gray
line) or $p=4$ (gray line). The values corresponding to $\sgx$,
$\sgf$ are also indicated. }\label{fig3}
\end{figure}

Practically the same observables for $n=0$ are shown in the right
plot of \Fref{fig1}. In that plot, though, we see that the models'
predictions are confined to $\ns\lesssim0.974$, $n$ may deviate
more appreciably from zero (mainly for $p=2$) and the maximal
possible $r$ is somewhat larger. Moreover, these predictions
depend harder on $\ca$ for $|n|>0.01$, as expected from
\eqs{ns2}{rs2}. Therefore, in that regime, we could say that these
models are less predictive than those based on $K=K_1$ and $K_3$.
Our results below are presented for $\ca$ such that $\sgx\simeq1$.
Namely, for $p=2$, we find
\beqs\bea\label{res25a} && 0.7\gtrsim{n}/{0.01}\gtrsim-3.5,\>\>\>
9.57\lesssim {\ns/0.1}\lesssim9.68,\\
\label{res25b} && 1.3\lesssim
{r}/{10^{-3}}\lesssim20\>\>\>\mbox{and}\>\>\>
0.98\lesssim{10^{5}\ld/\ca}\lesssim2.8\,, ~~~~~\eea\eeqs whereas
for $p=4$ we have
\beqs\bea\label{res45a} && 0.6\gtrsim{n}/{0.01}\gtrsim-1.45,\>\>\>
9.57\lesssim {\ns/0.1}\lesssim9.72,~~~~~\\
\label{res45b} &&2\lesssim
{r}/{10^{-3}}\lesssim14\>\>\>\mbox{and}\>\>\>
1.8\lesssim{10^{5}\ld/\ca^2}\lesssim3.6\,.~~~~~ \eea\eeqs
The latter results deviate more drastically from those in
\crefs{igHI,univ}. From the data of both plots of \Fref{fig1}, we
remark that $r\gtrsim0.0013$. These $r$ values are testable by the
forthcoming experiments \cite{bcp3}, which are expected to measure
$r$ with an accuracy of $10^{-3}$. The tuning, finally, required
for the attainment of hilltop nMI for $n>0$ is very low, since
$\sg_{\rm max}\gg\sgx$.

Although $\ld/\ca^{p/2}$ is constant for fixed $n$, the amplitudes
of $\ld$ and $\ca$ can be bounded. This fact is illustrated in
\Fref{fig2} where we display the allowed values $\ca$ versus
$\sqrt{\ld}$ for $p=4$ and $K=K_{1}$ (gray lines) or $K=K_5$
(light gray lines). We take $n=0$ (solid lines) and $n=-0.004$
(dashed lines). As anticipated in \Eref{camin} for any $n$ there
is a lower bound on $\ca$, above which $\sgx\leq1$ stabilizing
thereby the results against corrections from higher order terms --
e.g., $(\bar\Phi\Phi)^l$ with $l>1$ in \Eref{Whi}. The
perturbative bound $\ld=3.5$ limits the various lines at the other
end. We observe that the ranges of the allowed lines are much more
limited compared to other models -- cf. \crefs{nmH,nMkin} -- and
displaced to higher $\ld$ values as seen, also, by
\eqs{lan1}{lan2}. We find that for $p=5$, $\ld$ corresponding to
lowest possible $\ca$ violates the perturbative bound and so, our
proposal can not be applied for $p>4$.


\begin{figure}[!t]
\includegraphics[width=60mm,angle=-90]{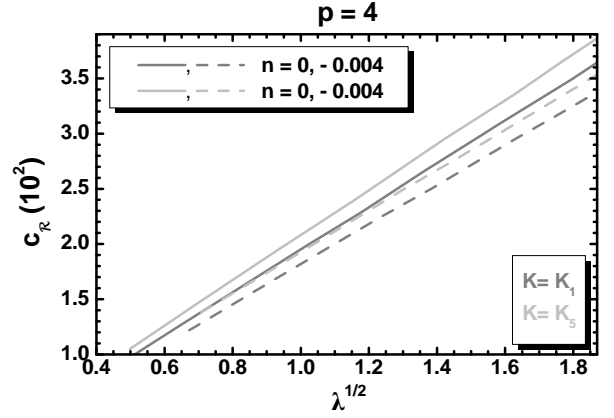}
\caption{\sl Allowed values of $\ca$ versus $\sqrt{\ld}$ for $p=4$
and various $n$'s and $K$'s. The conventions adopted for the
various lines are also shown.}\label{fig2}
\end{figure}


\section{A Post-Inflationary Completion}\label{post}

In a couple of recent papers \cite{univ,igHI} we attempt to
connect the high-scale inflationary scenario based on $W=\Whi$ in
\Eref{Whi} with the low energy physics, taking into account
constraints from the observed BAU, neutrino data and MSSM
phenomenology. It would be, therefore, interesting to check if
this scheme can be applied also in the case of our present set-up
where $\Whi$ in \Eref{Whi} cooperates with the $K$'s in
Eqs.~(\ref{K1}) -- (\ref{K5}) where $\fp$ and $\fm$ given in
\Eref{fpmH}. The necessary extra ingredients for such a scenario
are described in \Sref{post0}. Next, we show how we can correlate
nMI with the generation of the $\mu$ term of MSSM -- see
Sec.~\ref{post1} -- and the generation of BAU via nTL -- see
Sec.~\ref{post2}. Hereafter, we restore units, i.e., we take
$\mP=2.433\cdot10^{18}~\GeV$.


\renewcommand{\arraystretch}{1.45}
\begin{table*}[!t]
\caption{\normalfont Mass-squared spectrum of the non-inflaton
sector for various $K$'s along the path in \eqs{inftrn}{inftr1}.}
\begin{ruledtabular}
\begin{tabular}{c|c|c|c|c}
{\sc Fields}&{\sc Eigen-} & \multicolumn{3}{c}{\sc Masses
Squared}\\\cline{3-5}
&{\sc states} && {\hspace*{2cm}$K=K_1,K_2$\hspace*{2cm}}&{$K=K_3,K_4,K_5$} \\
\colrule
10 Real & $\widehat{h}_{\pm},\widehat{\bar h}_{\pm}$ &  $ \widehat
m_{h\pm}^2$&$3\Hhi^2\lf1+\ca\sg/{N}\pm{4\lm\fr/\ld\sg^2}\rg$
&{$3\Hhi^2\lf1+1/\nsu\pm{4\lm}/{\ld\sg^2}\rg$}\\\cline{3-5}
Scalars  & $\widehat{\tilde\nu}^c_{i},
\widehat{\bar{\tilde\nu}}^c_{i}$ & $\widehat m_{i\tilde
\nu^c}^2$&$3\Hhi^2\ca\lf \sg/N+16\ld^2_{iN^c}/{\ld^2\sg}\rg$

&{$3\Hhi^2\lf1+1/\nsu+16\ld^2_{iN^c }/\ld^2\sg^2\rg$}\\\colrule
$3$ Weyl Spinors &${\widehat N_i^c}$& $ \widehat
m_{{iN^c}}^2$&\multicolumn{2}{c}{$48\Hhi^2\ld^2_{iN^c
}/\ld^2\sg^2$}
\end{tabular}\label{tab3}
\end{ruledtabular}
\end{table*}
\renewcommand{\arraystretch}{1.}

\subsection{\sf\scshape\small Relevant Set-Up} \label{post0}

Following the post-inflationary setting of \cref{univ} we consider
a $B-L$ extension of MSSM with the field content charged under
$B-L$ and R as displayed in Table~1 therein. The superpotential of
the model contains $\Whi$ in \Eref{Whi}, the superpotential of
MSSM with $\mu=0$ and the following two terms
\beqs \bea \Wmu&=&\lm S\hu\hd\,,\label{Wmu} \\ \Wrhn
&=&\lrh[i]\phcb N^{c2}_i +h_{Nij} \sni L_j
\hu\,.\label{Wrhn}\eea\eeqs
From the terms above, the first one inspired by \cref{dvali} helps
to justify the existence of the $\mu$ term within MSSM, whereas
the second one allows for the implementation of (type I) see-saw
mechanism (providing masses to light neutrinos) and supports a
robust baryogenesis scenario through nTL. Let us note that $L_i$
denotes the $i$-th generation $SU(2)_{\rm L}$ doublet left-handed
lepton superfields, and $\hu$ [$\hd$] is the $SU(2)_{\rm L}$
doublet Higgs superfield which couples to the up [down] quark
superfields. Also, we assume that the superfields $N_j^c$ have
been rotated in the family space so that the coupling constants
$\ld_i$ are real and positive. This is the so-called
\cite{R2r,univ} $\sni$ basis, where the $\sni$ masses, $\mrh[i]$,
are diagonal, real and positive.

We assume that the extra scalar fields $X^\beta=\hu,\hd,\ssni$
have identical kinetic terms as the stabilizer field $S$ expressed
by the functions $F_{lS}$ with $l=1,2,3$ in Eqs.~(\ref{f1s}) --
(\ref{f3s}) -- see \cref{univ}. Therefore, $N_S$ may be renamed
$N_X$ henceforth. The inflationary trajectory in \Eref{inftr} has
to be supplemented by the conditions
\beq\label{inftr1} \hu=\hd=\ssni=0,\eeq
and the stability of this path has to be checked, parameterizing
the complex fields above as we do for $S$ in \Eref{cannor}. The
relevant masses squared are listed in \Tref{tab3} for $K=K_1 --
K_5$, with hatted fields being defined as $\widehat s$ and
$\widehat{\bar{s}}$ in \Eref{Jg}. Also we set
\beq \widehat h_\pm=(\widehat h_u\pm{\widehat
h_d})/\sqrt{2}\>\>\>\mbox{and}\>\>\> \widehat{\bar
h}_\pm=(\widehat{\bar h}_u\pm\widehat{\bar h}_d)/\sqrt{2}\,. \eeq
In \Tref{tab3} we see that $\what m^2_{i\tilde\nu^c}\gtrsim\Hhi^2$
and $\what m^2_{h+}\gtrsim\Hhi^2$ for every $\sg$, whereas
imposing $\what m^2_{h-}\gtrsim\Hhi^2$ dictates  \beq
\lm\lesssim\frac{\ld\sg^2}{4}\cdot\begin{cases} (2/3+\ca\sg/N)/\fr&\mbox{for}~~K=K_1, K_2,\\
(2/3+1/\nsu)&\mbox{for}~~K=K_3-K_5.\end{cases}\label{lm12}\eeq
Both bounds above depend on $\sg$ and assume their lowest values
for $\sg\simeq\sgf$. Taking, e.g., $n=0$ and $N_X=2$ the equations
above imply \beq
\label{lm12n}\lm\lesssim\begin{cases}1.5\cdot10^{-5}
&\mbox{for}~~K=K_1,
K_2,\\5.3\cdot10^{-5}&\mbox{for}~~K=K_3-K_5.\end{cases}\eeq
Since $\sgf$ is inverse proportional to $\ca$ and $\ld$ is
proportional to $\ca^2$ according to \eqs{sgf}{lan1} respectively,
the bounds above are practically independent of the variation of
$\ld$ and $\sgf$. Similar bounds are obtained in
\crefs{R2r,univ,igHI} and should not be characterized as
unnatural, given that the Yukawa coupling constant which provides
masses to the up-type quarks, is of the same order of magnitude at
a high scale -- cf.~\cref{fermionM}.

\subsection{\sf\scshape\small Solution to the {$\mu$} Problem of MSSM} \label{post1}

Supplementing, in \Sref{secmu2}, with soft SUSY breaking terms the
SUSY limit of the SUGRA potential -- found in \Sref{secmu1} -- we
can show that our model assists us to understand the origin of
$\mu$ term of MSSM, consistently with the low-energy phenomenology
-- see \Sref{pheno}.

\renewcommand{\arraystretch}{1.25}
\begin{table*}[!t]
\caption[]{\sl\small The required $\lm$ values rendering our
models for $n=0$ compatible with the best-fit points of the CMSSM
as found in \cref{mssm} with the assumptions in \Eref{softmass}.}
\label{tab}
\begin{ruledtabular}
\begin{tabular}{c|c|c|c||c|c|c|c|c}
\diagbox {\sc CMSSM}{\sc Parameters} &$|A_0|$&$m_0$&$|\mu|
$&$\am$&\multicolumn{4}{c}{\sc $\lm~(10^{-6})$}\\\cline{6-9} {\sc
Region
}&$(\TeV)$&$(\TeV)$&$(\TeV)$&&$K=K_1$&$K=K_2$&$K=K_3$&$K=K_4,~K_5$\\\colrule
$A/H$ Funnel &$9.9244$ &$9.136$&$1.409$&$1.086$ &$2.946$&$2.867$&$3.28$&$3.195$\\
$\tilde\tau_1-\chi$ Coannihilation &$1.2271$ &$1.476$&$2.62$& $0.831$&{\it 44.3}&{\it 43.13}&49.35&$48.06$\\
$\tilde t_1-\chi$ Coannihilation  &$9.965$ &$4.269$&$4.073$&$2.33$ &8.58&8.27&$9.461$&$9.214$\\
$\tilde \chi^\pm_1-\chi$ Coannihilation  &$9.2061$ &$9.000$&$0.983$&$1.023$ &$2.215$&$2.156$&$2.46$&$2.4$\\
\end{tabular}\label{tab4}
\end{ruledtabular}
\end{table*}\renewcommand{\arraystretch}{1.}

\subsubsection{\sf\small  SUSY Potential}\label{secmu1}

The presence of unity in the arguments of the logarithms of $K$'s
in Eqs.~(\ref{K1}) -- (\ref{K5}) with $\fp$ and $\fm$ defined in
\Eref{fpmH} allows us to expand them for $\mP\to\infty$ up to
quadratic terms obtaining $\widetilde K$. Focusing on the
$S-\phcb-\phc$ system we obtain
\beq \label{Kquad}\widetilde
K=-N\mP\ca\lf\fp+\fp^*\rg+\frac12\ca^2N \lf\fp+\fp^*\rg^2+
\fm+|S|^2\eeq
from which we can then compute
\beqs\beq \label{Kqab} \lf \widetilde K_{\al\bbet}\rg=\diag\lf
\widetilde M_\pm,1\rg \eeq where the matrix $\widetilde M_\pm$
reads
\beq \widetilde
M_\pm=\mtta{1+N\ca^2|\phcb|^2/4|\fp|^2}{N\ca^2\phcb\phc^*/4|\fp|^2}
{N\ca^2\phcb^*\phc/4|\fp|^2}{1+N\ca^2|\phc|^2/4|\fp|^2}\,.
\label{Mqpm}\eeq\eeqs
After calculating $\widetilde K^{-1}$ and substituting it into the
SUSY limit
\beq \label{Vsusy} V_{\rm SUSY}= \widetilde K^{\al\bbet} W_{\rm
HI\al} W^*_{\rm HI\bbet}+\frac{g^2}2 \mbox{$\sum_{\rm a}$} {\rm
D}_{\rm a} {\rm D}_{\rm a}\,,\eeq of $\Vhi$ in \Eref{Vsugra}, we
find
\bea \nonumber V_{\rm
SUSY}&=&\ld^2\left|\phcb\phc-\frac14{M^2}\right|^2
+\frac{\ld^2|S|^2}{\det\widetilde
M_\pm}\lf|\phc|^2+|\phcb|^2\rg\\&&+\frac{g^2}2\lf|\phc|^2-|\phcb|^2\rg^2,
\label{VF}\eea where the non-diagonal contributions in the F terms
proportional to $W_{\rm HI,\phc} W^*_{\rm HI,\phcb^*}$ and $W_{\rm
HI,\phcb} W^*_{\rm HI,\phc^*}$ are cancelled out and
\beq  \det\widetilde
M_\pm=1+\ca^2N\lf|\phc|^2+|\phcb|^2\rg/4|\fp|^2\,. \label{detMpm}
\eeq

From \Eref{VF}, we find that the SUSY vacuum lies along the D-flat
direction $|\phcb|=|\phc|$ with
\beq \vev{S}=0 \>\>\>\mbox{and}\>\>\>
|\vev{\Phi}|=|\vev{\bar\Phi}|=M/2\,.\label{vevs} \eeq
As a consequence, $\vev{\Phi}$ and $\vev{\bar\Phi}$ break
spontaneously $U(1)_{B-L}$ down to $\mathbb{Z}^{B-L}_2$. Since
$U(1)_{B-L}$ is already broken during nMI, no cosmic string are
formed. Given, finally, that $X^\bt$ participate in $\widetilde
K$, see \Eref{Kquad}, with terms similar to the one we have for
$S$, we can easily verify that their v.e.vs lie along the
direction in \Eref{inftr1}.

\subsubsection{\sf\small Generation of the $\mu$ Term of MSSM}\label{secmu2}

The contributions from the $\TeV$-scale soft SUSY-breaking terms,
although negligible during nMI, may shift slightly $\vev{S}$ from
zero in \Eref{vevs}. The relevant potential terms are
\beq V_{\rm soft}= \lf\ld A_\ld S \phcb\phc-{\rm a}_{S}S\ld M^2/4
+ {\rm h. c.}\rg+ m_{\al}^2\left|X^\al\right|^2, \label{Vsoft}
\eeq
where $m_{\al}$, $A_\ld$ and $\aS$ are soft SUSY-breaking mass
parameters. Considering $V_{\rm soft}$ together with $V_{\rm
SUSY}$ from \Eref{VF} we end up with the total low energy
potential \beq\label{Vtot} V_{\rm tot}=V_{\rm SUSY}+V_{\rm
soft}\,,\eeq which, replacing $\phc$ and $\phcb$ by their SUSY
v.e.vs from \Eref{vevs}, can be rephrased as
\beqs\beq \vev{V_{\rm tot}(S)}= \ld^2\,M^2S^2/(N\ca^2+2)-\ld\am
\mgr M^2 S, \label{Vol} \eeq
if we neglect $m_\al\ll M$ and set \beq\label{a32def} |A_\ld| +
|{\rm a}_{S}|=2\am\mgr\,.\eeq\eeqs Here $\mgr$ is the gravitino
($\Gr$) mass and $\am>0$ a parameter of order unity which
parameterizes our ignorance for the dependence of $|A_\ld|$ and
$|{\rm a}_{S}|$ on $\mgr$ -- note that the phases of $\Ald$ and
$\aS$ have been chosen so that $V_{\rm tot}$ is minimized and $S$
has been rotated in the real axis by an appropriate
$R$-transformation. The extremum condition for $\vev{V_{\rm
tot}(S)}$ in \Eref{Vol} w.r.t $S$ leads to a non-vanishing
$\vev{S}$ as follows
\beqs\beq \label{vevS}\frac{d}{d S} \vev{V_{\rm
tot}(S)}=0~~\Rightarrow~~\vev{S}\simeq \am\mgr(N\ca^2+2)/2\ld.\eeq
The generated $\mu$ term from $W_\mu$ in \Eref{Wmu} is \beq\mu
=\lm \vev{S} \simeq\lm\am\mgr(N\ca^2+2)/2\ld\,,\label{mu}\eeq\eeqs
which, although similar, is clearly distinguishable from the
results obtained in \crefs{R2r, univ, igHI}. The resulting $\mu$
above depends only on $n$ and $\lm$ since $\ld/\ca^2$ is fixed for
frozen $n$ by virtue of \Erefs{lan1}{lan3}. As a consequence, we
may verify that any $|\mu|$ value is accessible for the $\lm$'s
allowed by \Eref{lm12n} without any ugly hierarchy between $\mgr$
and $\mu$.

\subsubsection{\sf\small Link to the MSSM Phenomenology}\label{pheno}

The subgroup, $\mathbb{Z}_2^{R}$ of $U(1)_R$ -- which remains
unbroken after the consideration of the SUSY breaking effects in
\Eref{Vsoft} -- combined with the $\mathbb{Z}_2^{\rm f}$ fermion
parity yields the well-known $R$-parity. This symmetry guarantees
the stability of the \emph{lightest SUSY particle} ({\sf\ftn
LSP}), providing thereby a well-motivated \emph{cold dark matter}
({\sf\ftn CDM}) candidate.

The candidacy of LSP may be successful, if it generates the
correct CDM abundance \cite{plcp} within a concrete low-energy
framework. In the case under consideration \cite{univ} this could
be the \emph{Constrained MSSM} ({\ftn\sf CMSSM}), which employs
the following free parameters
\begin{equation}
{\rm
sign}\mu,~~\tan\beta=\vev{\hu}/\vev{\hd},~~\mg,~~m_0~~\mbox{and}~~A_0,
\label{para}
\end{equation}
where ${\rm sign}\mu$ is the sign of $\mu$, and the three last
mass parameters denote the common gaugino mass, scalar mass and
trilinear coupling constant, respectively, defined (normally) at
$\Mgut$. The parameter $|\mu|$ is not free, since it is computed
at low scale by enforcing the conditions for the electroweak
symmetry breaking. The values of the parameters in \Eref{para} can
be tightly restricted imposing a number of cosmo-phenomenological
constraints from which the consistency of LSP relic density with
observations plays a central role. Some updated results are
recently presented in \cref{mssm}, where we can also find the
best-fit values of $|A_0|$, $m_0$ and $|\mu|$ listed in the
second, third and fourth leftmost columns of \Tref{tab4}.  We see
that there are four allowed regions characterized by the specific
mechanism for suppressing the relic density of the LSP which is
the lightest neutralino ($\chi$) -- note that $\tilde\tau_1,
\tilde t_1$ and $\tilde \chi^\pm_1$ stand for the lightest stau,
stop and chargino eigenstate.

The inputs from \cref{mssm} can be deployed within our setting, if
we identify, e.g.,
\beq
m_0=\mgr~~\mbox{and}~~|A_0|=|A_\ld|=|\aS|.\label{softmass}\eeq
Fixing also $n=0$, we first derive $\am$ from \Eref{a32def} -- see
fifth column of \Tref{tab4} -- and then the $\lm$ values which
yield the phenomenologically desired $|\mu|$ -- ignoring
renormalization group effects. The outputs w.r.t $\lm$ of our
computation are listed in the four rightmost columns of
\Tref{tab4} for $K=K_1 - K_5$. From these we infer that the
required $\lm$ values vary slightly depending on $N$ and $\ca$
required by each $K$ and, besides the ones written in italics, are
comfortably compatible with \Eref{lm12n}. Therefore, the whole
inflationary scenario can be successfully combined with all the
allowed regions CMSSM besides the $\tilde\tau_1-\chi$
coannihilation region for $K=K_1$ and $K_2$. On the other hand,
the $\mgr$'s used in all the CMSSM regions can be consistent with
the gravitino limit on reheat temperature $\Trh$, under the
assumption of the unstable $\Gr$, for the $\Trh$ values
necessitated for satisfactory leptogenesis-- see \Sref{lept1}.


\subsection{\sf\scshape\small Non-Thermal Leptogenesis and Neutrino
Masses}\label{post2}

Besides the generation of $\mu$ term, our post-inflationary
setting offers a graceful exit from the inflationary phase (see
\Sref{lept0}) and explains the observed BAU (see \Sref{lept1})
consistently with the $\Gr$ constraint and the low energy neutrino
data. Our results are summarized in \Sref{num}.

\subsubsection{\sf\small  Inflaton Mass \& Decay}\label{lept0}

When nMI is over, the inflaton continues to roll down towards the
SUSY vacuum, \Eref{vevs}. Soon after, it settles into a phase of
damped oscillations around the minimum of $\Vhi$. The (canonically
normalized) inflaton,
\beq\dphi=\vev{J}\dph\>\>\>\mbox{with}\>\>\> \dph=\phi-M,
\label{dphi}\eeq where $\vev{J}$ is estimated from \Eref{vevJ},
acquires mass given by
\beq \label{msn} \msn=\left\langle\Ve_{\rm
HI,\se\se}\right\rangle^{1/2}=\frac{\ld
M}{\sqrt{2\vev{J}}}\simeq\frac{\ld M}{\sqrt{N}\ca}.\eeq
As we see, $\msn$ depends crucially on $M$ which is bounded from
above by the requirement $\vev{\fr}=1$ ensuring the establishment
of the conventional Einstein gravity at the vacuum. This bound is
translated to an upper bound on the mass $\vev{M_{BL}}$ that the
$B-L$ gauge boson acquires for $\sg=\vev{\sg}$ -- see \Tref{tab2}.
Namely we obtain $\vev{M_{BL}}\leq10^{14}~\GeV$, which is lower
than the value $\Mgut\simeq2\cdot10^{16}~\GeV$ dictated by the
unification of the MSSM gauge coupling constants -- cf.
\crefs{univ, igHI}. However, since $U(1)_{B-L}$ gauge symmetry
does not disturb this unification, we can treat $\vev{M_{BL}}=gM$
as a free parameter with $g\simeq0.5-0.7$ being the value of the
GUT gauge coupling at the scale $\vev{M_{BL}}$.

During the phase of its oscillations at the SUSY vacuum, $\dphi$
decays perturbatively  reheating the Universe at a reheat
temperature given by \cite{rh}
\beqs\beq \label{T1rh} \Trh= \left({40/\pi^2g_{\rm
rh*}}\right)^{1/4}\Gsn^{1/2}\mP^{1/2},\eeq where the unusual --
cf. \crefs{univ,igHI} -- prefactor is consistent with $w_{\rm
rh}\simeq0.33$ \cite{rh} and we set $g_{\rm rh*}=228.75$ as in
\Eref{Prob}. The total decay width of $\dphi$ is found to be
\beq\Gsn=\GNsn+\Ghsn+\Gysn\,,\label{Gol}\eeq\eeqs
where the individual decay widths are
\beqs\bea \GNsn&=&\frac{g_{iN^c}^2}{16\pi}\msn\lf1-\frac{4\mrh[
i]^2}{\msn^2}\rg^{3/2};\\
\Ghsn&=&\frac{2}{8\pi}g_{H}^2\msn;\\
\Gysn&=&g_y^2\frac{7}{256\pi^3}\frac{\msn^3}{\mP^2},\eea\eeqs
and the relevant coupling constants are defined as follows
\beqs\bea && g_{iN^c}=\frac{\ld_{iN^c}}{\vev{J}}\lf1-N\ca
\frac{M}{\mP}\rg,\\ && g_{H}\simeq{\lm}/{\sqrt{2}}~~\mbox{and}~~
g_y=y_{3}N\vev{J}\ca/{2}\,.\eea\eeqs
The decay widths above arise from the lagrangian terms
\beqs\bea \nonumber {\cal L}_{\dphi\to
\sni\sni}&=&-\frac12e^{K/2\mP^2}W_{{\rm RHN},N_i^cN^c_i}\sni\sni\
+\ {\rm h.c.}~~~~~~\\&=&g_{iN^c} \dphi\ \lf\sni\sni\ +\ {\rm
h.c.}\rg +\cdots;\\\nonumber {\cal
L}_{\dphi\to\hu\hd}&=&-e^{K/\mP^2}K^{SS^*}\left|W_{\mu,S}\right|^2
\\&=&-g_{H} \msn\dphi \lf H_u^*H_d^*\ +\ {\rm
h.c.}\rg+\cdots;~~~\\ \nonumber{\cal L}_{\dphi\to
X\psi_Y\psi_Z}&=&-\frac12e^{{K/2\mP^2}}\lf
W_{y,YZ}\psi_{Y}\psi_{Z}\rg+{\rm h.c.}\,\\
&=&-g_y\frac\dphi\mP\lf X\psi_{Y}\psi_{Z}\rg+{\rm
h.c.},\label{lxyz} \eea\eeqs
describing respectively $\dphi$ decay into a pair of $N^c_{j}$
with masses $\mrh[j]=\ld_{jN^c}M<\msn/2$, $\hu$ and $\hd$ and
three MSSM (s)-particles $X,Y,Z$ involved in a typical trilinear
superpotential term $W_y=yXYZ$. Here, $\psi_X, \psi_{Y}$ and
$\psi_{Z}$ are the chiral fermions associated with the superfields
$X, Y$ and $Z$ whose scalar components are denoted with the
superfield symbols and $y=y_3\simeq(0.4-0.6)$ is a Yukawa coupling
constant of the third generation.

\subsubsection{\sf\small Lepton-Number and Gravitino Abundances}\label{lept1}

For $\Trh<\mrh[i]$, the out-of-equilibrium decay of $N^c_{i}$
generates a lepton-number asymmetry (per $N^c_{i}$ decay), $\ve_i$
estimated from \cref{plum}. The resulting lepton-number asymmetry
is partially converted through sphaleron effects into a yield of
the observed BAU
\beq Y_B=-0.35\frac{3}{2}{\Trh\over\msn}\mbox{$\sum_i$}
{\GNsn\over\Gsn}\ve_i,\label{Yb}\eeq
where $\vev{\hu}\simeq174~\GeV$, for large $\tan\beta$ and $m_{\rm
D}$ is the Dirac mass matrix of neutrinos, $\nu_i$. The ratio
($3/2$) is again \cite{rh} consistent with $w_{\rm rh}=0.33$. The
expression above has to reproduce the observational result
\cite{plcp}
\beq
Y_B=\lf8.64^{+0.15}_{-0.16}\rg\cdot10^{-11}.\label{BAUwmap}\eeq
The validity of \Eref{Yb} requires that the $\dphi$ decay into a
pair of $\sni$'s is kinematically allowed for at least one species
of the $\sni$'s and also that there is no erasure of the produced
$Y_L$ due to $N^c_1$ mediated inverse decays and $\Delta L=1$
scatterings. These prerequisites are ensured if we impose
\beq\label{kin} {\sf \ftn
(a)}\>\>\msn\geq2\mrh[1]\>\>\>\mbox{and}\>\>\>{\sf \ftn
(b)}\>\>\mrh[1]\gtrsim 10\Trh.\eeq
The quantity $\ve_i$ can be expressed in terms of the Dirac masses
of $\nu_i$, $\mD[i]$, arising from the third term of \Eref{Whi}.
Employing the (type I) seesaw formula we can then obtain the
light-neutrino mass matrix $m_\nu$ in terms of $\mD[i]$ and
$\mrh[i]$. As a consequence, nTL can be nicely linked to low
energy neutrino data. We take as inputs the recently updated
best-fit values \cite{valle} -- cf. \cref{univ} -- on the neutrino
mass-squared differences, $\Delta m^2_{21}=7.56\cdot10^{-5}~{\rm
eV}^2$ and $\Delta m^2_{31}=2.55\cdot10^{-3}~{\rm eV}^2$
$\left[\Delta m^2_{31}=2.49\cdot10^{-3}~{\rm eV}^2\right]$, on the
mixing angles, $\sin^2\theta_{12}=0.321$,
$\sin^2\theta_{13}=0.02155~\left[\sin^2\theta_{13}=0.0214\right]$
and $\sin^2\theta_{23}=0.43~\left[\sin^2\theta_{23}=0.596\right]$
and the CP-violating Dirac phase
$\delta=1.4\pi~\left[\delta=1.44\pi\right]$ for \emph{normal
[inverted] ordered} ({\ftn\sf NO [IO]}) \emph{neutrino masses},
$\mn[i]$'s. Furthermore, the sum of $\mn[i]$'s is bounded from
above at 95\% c.l. by the data \cite{plcp} \beq\mbox{$\sum_i$}
\mn[i]\leq0.23~{\eV}.\label{sumnu}\eeq

The required $\Trh$ in \Eref{Yb} must be compatible with
constraints on the $\Gr$ abundance, $Y_{3/2}$, at the onset of
\emph{nucleosynthesis} (BBN), which is estimated to be
\cite{brand,kohri}
\beq\label{Ygr} Y_{3/2}\simeq 1.9\cdot10^{-22}\ \Trh/\GeV,\eeq
where we take into account only thermal production of $\Gr$, and
assume that $\Gr$ is much heavier than the MSSM gauginos.
Non-thermal contributions to $\Yg$ \cite{Idecay} are also possible
but strongly dependent on the mechanism of soft SUSY breaking. No
precise computation of this contribution exists within nMI
adopting the simplest Polonyi model of SUSY breaking \cite{grNew}.
It is notable, though, that the non-thermal contribution to $\Yg$
in models with stabilizer field, as in our case, is significantly
suppressed compared to the thermal one.

On the other hand, $\Yg$  is bounded from above in order to avoid
spoiling the success of the BBN. For the typical case where $\Gr$
decays with a tiny hadronic branching ratio, we obtain
\cite{kohri} an upper bound on $\Trh$, i.e.,
\beq  \label{Ygw} \Trh\lesssim5.3\cdot\left\{\bem
%
10^{7}~\GeV\hfill \cr
10^{8}~\GeV\hfill \cr
10^{9}~\GeV\hfill \cr\eem
\right.\>\>\>\mbox{for}\>\>\>\mgr\simeq\left\{\bem
0.69~\TeV\,,\cr
10.6~\TeV\,,\cr
13.5~\TeV\,.\cr \eem
\right.\eeq
The bounds above can be somehow relaxed in the case of a stable
$\Gr$.

\subsubsection{\sf\small Results}\label{num}

Confronting with observations $Y_B$ and $\Trh$ which depend on
$\msn$, $\mrh[i]$ and $\mD[i]$'s  -- see \eqs{Yb}{Ygw} -- we can
further constrain the parameter space of the our models. In our
investigation we follow the bottom-up approach detailed in
\cref{univ}, according to which we find the $\mrh[i]$'s by using
as inputs the $\mD[i]$'s, a reference mass of the $\nu_i$'s --
$\mn[1]$ for NO $\mn[i]$'s, or $\mn[3]$ for IO $\mn[i]$'s --, the
two Majorana phases $\varphi_1$ and $\varphi_2$ of the PMNS
matrix, and the best-fit values for the low energy parameters of
neutrino physics mentioned in \Sref{lept1}. In our numerical code,
we also estimate \cite{univ} the renormalization-group evolved
values of the latter parameters at the scale of nTL,
$\Lambda_L=\msn$, by considering the MSSM with $\tan\beta\simeq50$
as an effective theory between $\Lambda_L$ and the soft SUSY
breaking scale, $M_{\rm SUSY}=1.5~\TeV$. We evaluate the
$\mrh[i]$'s at $\Lambda_L$, and we neglect any possible running of
the $\mD[i]$'s and $\mrh[i]$'s. The so obtained $\mrh[i]$'s
clearly correspond to the scale $\Lambda_L$.

\renewcommand{\arraystretch}{1.25}
\begin{table}[!t]
\caption{\sl Parameters yielding the correct BAU for $K=K_1$ or
$K_2$, $\vev{M_{BL}}=10^{12}~\GeV$, $n=0$, $\lm=10^{-6}$,
$y_3=0.5$ and various neutrino mass schemes.}
\begin{ruledtabular}\begin{tabular}{c||c|c||c|c|c||c|c}
{\sc Cases}& A&B& C & D& E & F&G\\
\cline{2-8} \diagbox{\sc }{\sc $~~~~~~~~$}
&\multicolumn{2}{c||}{Normal} & \multicolumn{3}{c||}{Almost}&
\multicolumn{2}{c}{Inverted}
\\{\sc Parameters}& \multicolumn{2}{c||}{Hierarchy}&\multicolumn{3}{c||}{Degeneracy}& \multicolumn{2}{c}{Hierarchy}\\
\colrule
\multicolumn{8}{c}{Low Scale Parameters (Masses in
$\eV$)}\\\colrule
$\mn[1]/0.1$&$0.01$&$0.1$&$0.5$ & $0.6$& $0.7$ & $0.51$&$0.5$\\
$\mn[2]/0.1$&$0.09$&$0.13$&$0.51$ & $0.61$& $0.71$ & $0.52$&$0.51$\\
$\mn[3]/0.1$&$0.5$&$0.51$&$0.71$ & $0.78$&$0.5$ &
$0.1$&$0.05$\\\colrule
$\sum_i\mn[i]/0.1$&$0.6$&$0.74$&$1.7$ & $2$&$1.9$ & $1.1$&$1$\\
%
%
\colrule
$\varphi_1$&$-\pi/9$&$-2\pi/3$&$\pi$ & $\pi/2$&$\pi/2$ & $3\pi/5$&$2\pi/3$\\
$-\varphi_2$&$\pi/2$&$0$ &$-\pi/2$& $2\pi/3$&$2\pi/3$ &
$\pi/2$&$\pi/2$\\\colrule
\multicolumn{8}{c}{Leptogenesis-Scale Mass Parameters in
$\GeV$}\\\colrule
$\mD[1]/0.1$&$0.38$&$0.69$&$3.23$ & $2.25$&$0.94$ & $50$&$1.9$\\
$\mD[2]$&$1$&$5$&$0.2$ & $0.15$&$0.3$ & $0.097$&$0.2$\\
$\mD[3]$&$1$&$10$&$5$ & $10$&$9$ & $6$&$5$\\\colrule
$\mrh[1]/10^{8}$&$5.4$&$8.4$&$9.9$ & $4.7$&$2.2$ & $7.9$&$12.6$\\
$\mrh[2]/10^{9}$& $13$&$755$&$2.5$&$0.96$ &$1.57$ & $219$&$3.5$\\
$\mrh[3]/10^{11}$&$3.2$&$18.8$&$1.6$ & $6.1$&$5$ &
$12.7$&$12.1$\\\colrule
\multicolumn{8}{c}{Decay channels of $\dphi$}\\\colrule
$\dphi\to$&$\wrhn[1]$&$\wrhn[1]$& $\wrhn[1]$& $\wrhn[1,2]$&
$\wrhn[1,2]$ & $\wrhn[1]$&$\wrhn[1]$\\
%
\colrule
\multicolumn{8}{c}{Resulting $B$-Yield }\\\colrule
%
%
$Y_B/10^{-11}$&$8.7$&$8.6$& $8.6$& $8.6$&$8.5$ &
$8.7$&$8.6$\\\colrule
\multicolumn{8}{c}{Resulting $\Trh$ (in $\GeV$)}\\\colrule
$\Trh/10^{7}$&$2.8$&$3$& $3$& $3.1$&$2.7$ & $2.9$&$3.3$\\
%
\end{tabular}
\end{ruledtabular} \label{tab5}
\end{table}
\renewcommand{\arraystretch}{1.}

\begin{figure*}[!t]
\includegraphics[width=60mm,angle=-90]{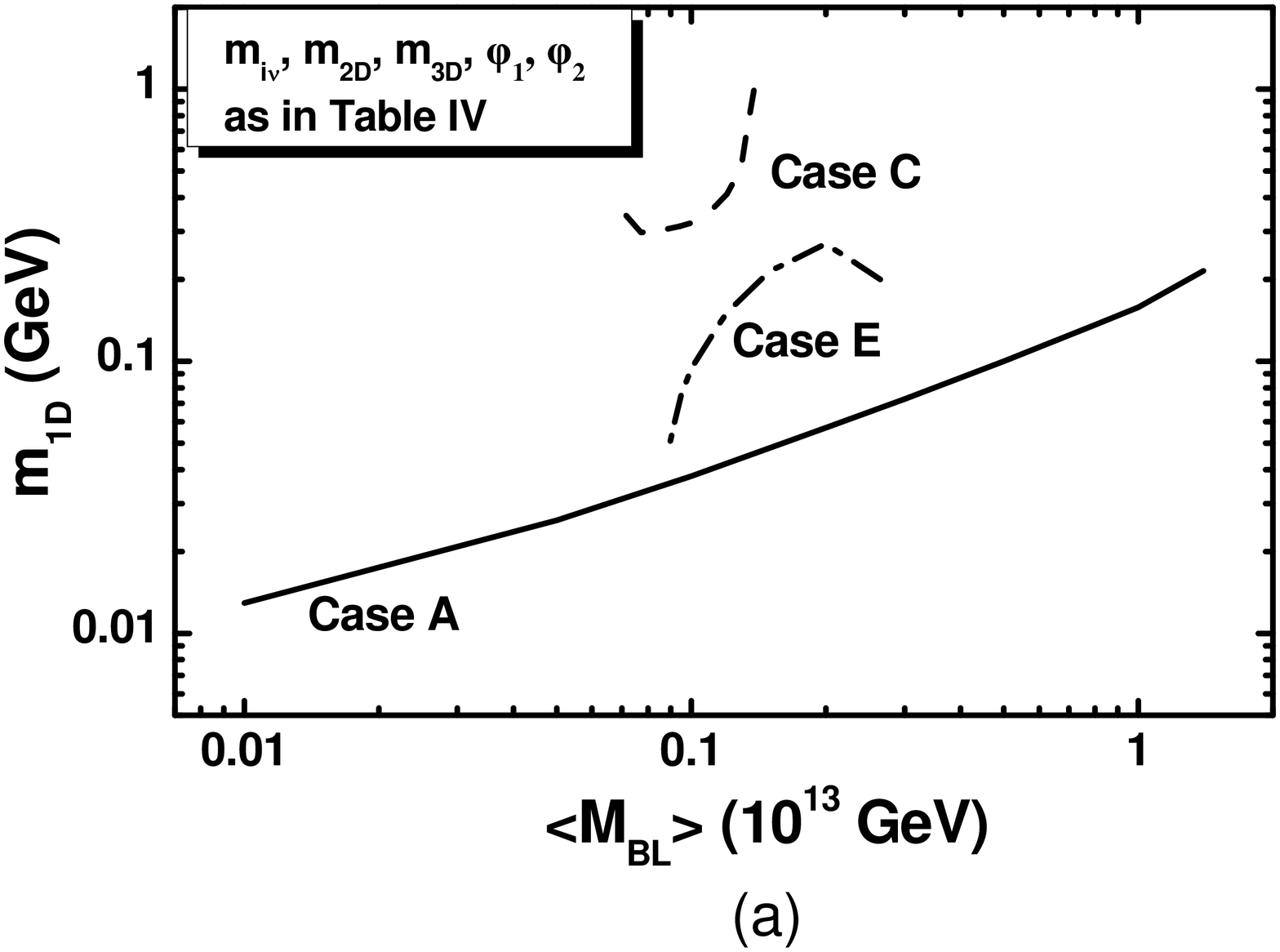}
\includegraphics[width=60mm,angle=-90]{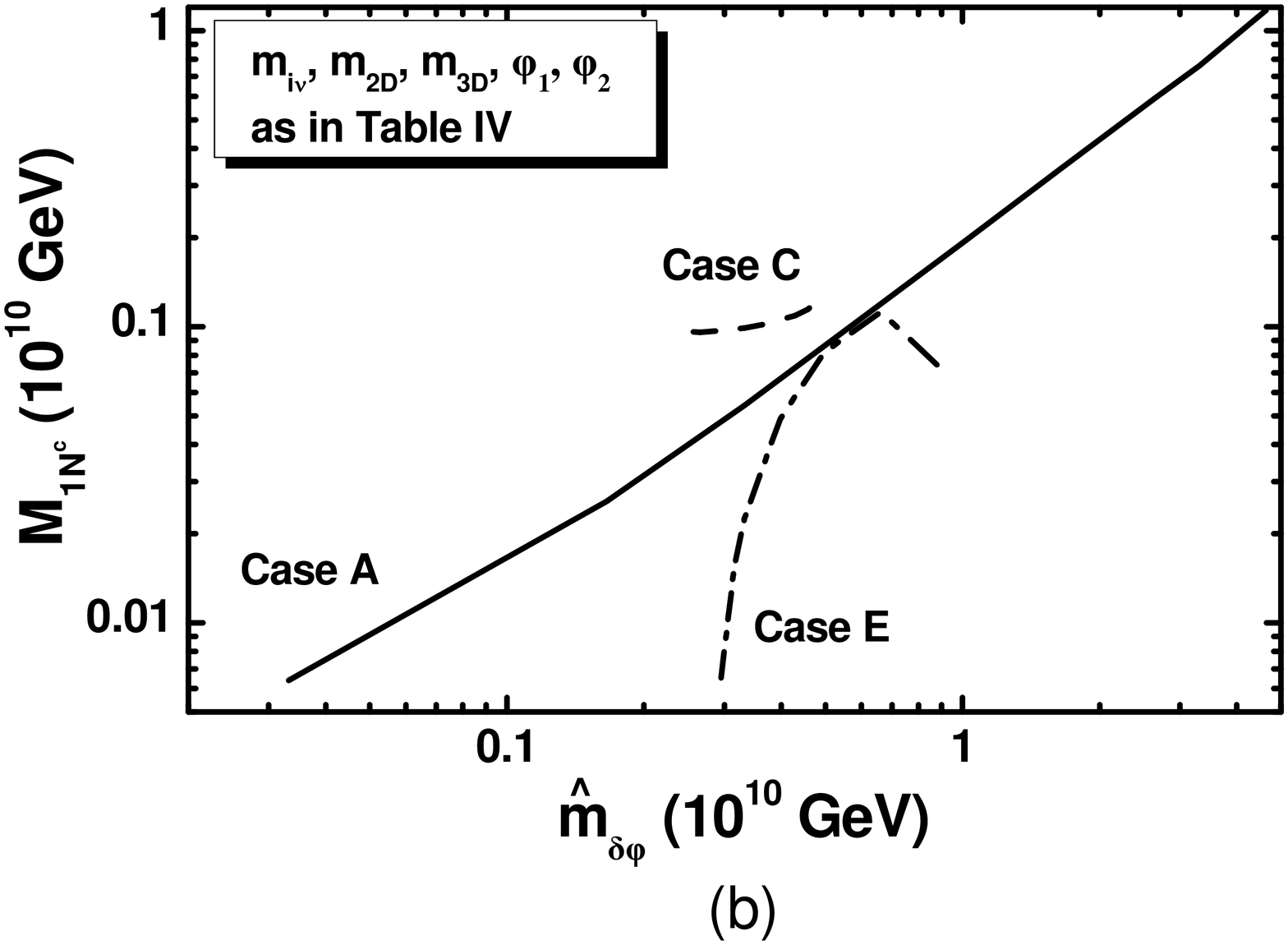}
\caption{\sl  Contours, yielding the central $Y_B$ in Eq.~(5.8)
consistently with the inflationary requirements, in the
{\sffamily\ftn (a)} $\vev{M_{BL}}-m_{\rm 1D}$ and {\sffamily\ftn
(b)} $\widehat{m}_{\delta\phi}-M_{1N^c}$ plane. We take $K=K_1$,
$n=0$, $\lm=10^{-6}$ and the values of $m_{i\nu}$, $m_{\rm 2D}$,
$m_{\rm 3D}$, $\varphi_1$ and $\varphi_2$ which correspond to the
cases A (solid line), C (dashed line) and E (dot-dashed line) of
\Tref{tab5}.}\label{fmD}
\end{figure*}

Some representative values of the parameters which yield $\Yb$ and
$\Trh$ compatible with \eqs{BAUwmap}{Ygw}, respectively, are
arranged in \Tref{tab5}. We take $n=0$ -- to avoid any tuning as
regards the inflationary inputs --, $\lm=10^{-6}$ in accordance
with \Eref{lm12n}, and $\vev{M_{BL}}=10^{12}~\GeV$. Note that we
consider $\vev{M_{BL}}$ as a free parameter since the unification
value -- imposed in \crefs{univ,igHI} -- is not reconciled with
the reappearance of Einstein gravity at low energies, i.e.,
$\vev{\fr}=1$. Setting $g=0.7$ in the formula giving $M_{BL}$ in
\Tref{tab2}, we obtain $M=1.43\cdot10^{12}~\GeV$ resulting via
\Eref{msn} to $2.8\leq\msn/10^{9}~\GeV\leq4.1$ -- the variation is
due to the choice of $K$. Although this amount of uncertainty does
not cause any essential alteration of the final outputs, we
mention just for definiteness that we take throughout $K=K_1$
corresponding to $\msn=3.3\cdot10^{10}~\GeV$. We consider NO
(cases A and B), almost degenerate (cases C, D and E) and IO
(cases F and G) $\mn[i]$'s. In all cases, the current limit in
\Eref{sumnu} is safely met.  This is more restrictive than the
$90\%$ c.l. upper bound arising from the effective electron
neutrino mass $m_{\beta}$ in $\beta$-decay \cite{2beta} by various
experiments. Indeed, the current upper bounds on $m_{\beta}$ are
comfortably satisfied by the values found in our set-up \beq
\label{2beta} 0.002\leq m_{\beta}/\eV\leq0.036,\eeq where the
lower and upper bound corresponds to case A and C respectively.

The gauge symmetry considered here does not predict any particular
Yukawa unification pattern and so, the $\mD[i]$'s are free
parameters. This fact allows us to consider $\mD[i]$'s which are
not hierarchical depending on the generation. Also, it facilitates
the fulfilment of \sEref{kin}{b} since $\mD[1]$ affects heavily
$\mrh[1]$. Care is also taken so that the perturbativity of
$\ld_{iN^c}$ -- defined below \Eref{lxyz} -- holds, i.e.,
$\ld_{iN^c}^2/4\pi\leq1$. The inflaton $\dphi$ decays mostly into
$N_1^c$'s -- see cases A -- E. In all cases $\GNsn<\Ghsn$ and so
the ratios $\GNsn/\Gsn$ introduce a considerable reduction in the
derivation of $\Yb$. Namely, we obtain
\beq0.07\lesssim\GNsn/\Gsn\lesssim0.35\eeq where the lower [upper]
bound comes out in case E [G]. In \Tref{tab5} we also display the
values of $\Trh$, the majority of which are close to
$3\cdot10^7~\GeV$, and consistent with \Eref{Ygw} for
$\mgr\gtrsim1~\TeV$. These values are in nice agreement with the
ones needed for the solution of the $\mu$ problem of MSSM -- see,
e.g., \Tref{tab4}. Thanks to our non-thermal set-up, successful
leptogenesis can be accommodated with $\Trh$'s lower than those
necessitated in the thermal regime -- cf. \cref{stefan}.

In order to investigate the robustness of the conclusions inferred
from \Tref{tab5}, we examine also how the central value of $Y_B$
in \Eref{BAUwmap} can be achieved by varying $\vev{M_{BL}}$, or
$\msn$, and adjusting conveniently $\mD[1]$ or $\mrh[1]$ -- see
\sFref{fmD}{a} and ({\ftn\sf b}) respectively. We fix again $n=0$
and $\lm=10^{-6}$. Since the range of $Y_B$ in \Eref{BAUwmap} is
very narrow, the $95\%$ c.l. width of these contours is
negligible. The convention adopted for the various lines is also
depicted. In particular, we use solid, dashed and dot-dashed line
when the remaining inputs -- i.e. $\mn[i]$, $\mD[2]$, $\mD[3]$,
$\varphi_1$, and $\varphi_2$ -- correspond to the cases A, C and E
of \Tref{tab5}, respectively. At the lower limit of these lines
nTL becomes inefficient (due to low $\Trh$) failing to reach the
value in \Eref{BAUwmap}. At the other end, these lines terminate
at the values of $m_{\rm 1D}$ beyond which \sEref{kin}{b} is
violated and, therefore, washout effects start becoming
significant. Along the depicted contours, the resulting $\mrh[2]$
and $\mrh[3]$ remain close to their values presented in the
corresponding cases of \Tref{tab4}. As regards the other
quantities, in all we obtain
\beqs\bea && 0.04\lesssim\Trh/{10^{8}~\GeV}\lesssim13,\\
&& 0.03\lesssim\msn/10^{10}~\GeV\lesssim4.64,\label{res3}\eea\eeqs
with the lower and upper bound obtained in the limits of the solid
line which represent the most ample region of parameters
satisfying the imposed requirements. These values are much lower
than those obtained in \cref{igHI} and a little lower than those
found in \cref{univ}, mainly due to lower $\vev{M_{BL}}$'s
employed here.

As a bottom line, nTL is a realistic possibility within our
setting. It can be comfortably reconciled with the $\Gr$
constraint even for $\mgr\sim1~\TeV$ as deduced from
\eqs{res3}{Ygw} adopting a sufficiently low $\vev{M_{BL}}$.

\section{Conclusions}\label{con}

Motivated by the fact that a strong linear non-minimal coupling of
the inflaton to gravity does not cause any problem with the
validity of the effective theory up to the Planck scale, we
explored the possibility to attain observationally viable nMI
(i.e. non-minimal inflation) in the context of standard SUGRA by
strictly employing this coupling. We showed that nMI is easily
achieved, for $p\leq4$ in the superpotential of \Eref{Wci}, by
conveniently adjusting the prefactor $(-N)$ of the logarithmic
part of the relevant \Ka s $K$ given in Eqs.~(\ref{K1}) --
(\ref{K5}), where the relevant functions $\fp$ and $\fm$ are shown
in \Eref{fpm} for a gauge-singlet inflaton. For appropriately
selected integer $N$'s -- i.e., setting $n=0$ in \Eref{ndef} --,
our models retain the predictive power of well-known universal
attractor models -- which employ a non-minimal coupling
functionally related to the potential -- and yield similar
results. Allowing for non-integer $N$ values, this predictability
is lost since the observables depend on the adopted $K$ and $n$ in
\Eref{ndef} and may yield any $\ns$ in its allowed region and
$0.0013\leq r\leq0.02$.

This scheme works also for a gauge non-singlet inflaton employing
the superpotential shown in \Eref{Whi} and the functions $\fp$ and
$\fm$ in \Eref{fpmH}. Embedding these models within a $B-L$
extension of MSSM, we showed that a $\mu$ term is easily generated
and the baryon asymmetry in the Universe is naturally explained
via non-thermal leptogenesis. The $B-L$ breaking scale
$\vev{M_{BL}}$, though, has to take values lower than the MSSM
unification scale and so, the present scheme is similarly
predictive with that of \cref{univ} which employs one more
parameter in the \Ka s but allows for $\vev{M_{BL}}$'s fixed by
the gauge coupling unification within MSSM. Our scenario can be
comfortably tolerated with almost all the allowed regions of the
CMSSM with gravitino as low as $1~\TeV$. Moreover, leptogenesis is
realized through the out-of equilibrium decay of the inflaton to
the right-handed neutrinos $N_1^c$ and/or $N_2^c$ with masses
lower than $2.32\cdot10^{10}~\GeV$, and a reheat temperature
$\Trh\leq10^9~\GeV$ taking $\vev{M_{BL}}\leq10^{13}~\GeV$.

\paragraph*{\small\bfseries\scshape Acknowledgments} {\small I would like to acknowledge
A. Riotto for a useful discussion and encouragement.}


\def\ijmp#1#2#3{{\sl Int. Jour. Mod. Phys.}
{\bf #1},~#3~(#2)}
\def\plb#1#2#3{{\sl Phys. Lett. B }{\bf #1}, #3 (#2)}
\def\prl#1#2#3{{\sl Phys. Rev. Lett.}
{\bf #1},~#3~(#2)}
\def\rmp#1#2#3{{Rev. Mod. Phys.}
{\bf #1},~#3~(#2)}
\def\prep#1#2#3{{\sl Phys. Rep. }{\bf #1}, #3 (#2)}
\def\prd#1#2#3{{\sl Phys. Rev. D }{\bf #1}, #3 (#2)}
\def\npb#1#2#3{{\sl Nucl. Phys. }{\bf B#1}, #3 (#2)}
\def\npps#1#2#3{{Nucl. Phys. B (Proc. Sup.)}
{\bf #1},~#3~(#2)}
\def\mpl#1#2#3{{Mod. Phys. Lett.}
{\bf #1},~#3~(#2)}
\def\jetp#1#2#3{{JETP Lett. }{\bf #1}, #3 (#2)}
\def\app#1#2#3{{Acta Phys. Polon.}
{\bf #1},~#3~(#2)}
\def\ptp#1#2#3{{Prog. Theor. Phys.}
{\bf #1},~#3~(#2)}
\def\n#1#2#3{{Nature }{\bf #1},~#3~(#2)}
\def\apj#1#2#3{{Astrophys. J.}
{\bf #1},~#3~(#2)}
\def\mnras#1#2#3{{MNRAS }{\bf #1},~#3~(#2)}
\def\grg#1#2#3{{Gen. Rel. Grav.}
{\bf #1},~#3~(#2)}
\def\s#1#2#3{{Science }{\bf #1},~#3~(#2)}
\def\ibid#1#2#3{{\it ibid. }{\bf #1},~#3~(#2)}
\def\cpc#1#2#3{{Comput. Phys. Commun.}
{\bf #1},~#3~(#2)}
\def\astp#1#2#3{{Astropart. Phys.}
{\bf #1},~#3~(#2)}
\def\epjc#1#2#3{{Eur. Phys. J. C}
{\bf #1},~#3~(#2)}
\def\jhep#1#2#3{{\sl J. High Energy Phys.}
{\bf #1}, #3 (#2)}
\newcommand\jcap[3]{{\sl J.\ Cosmol.\ Astropart.\ Phys.\ }{\bf #1}, #3 (#2)}
\newcommand\njp[3]{{\sl New.\ J.\ Phys.\ }{\bf #1}, #3 (#2)}
\def\prdn#1#2#3#4{{\sl Phys. Rev. D }{\bf #1}, no. #4, #3 (#2)}
\def\jcapn#1#2#3#4{{\sl J. Cosmol. Astropart.
Phys. }{\bf #1}, no. #4, #3 (#2)}
\def\epjcn#1#2#3#4{{\sl Eur. Phys. J. C }{\bf #1}, no. #4, #3 (#2)}

\rhead[\fancyplain{}{ \bf \thepage}]{\fancyplain{}{\sl
Unitarity-Safe Models of nMI in SUGRA}} \lhead[\fancyplain{}{\sl
References}]{\fancyplain{}{\bf \thepage}} \cfoot{}

\end{document}